\documentclass[twocolumn,prb]{revtex4}

\usepackage{graphicx}
\usepackage{epsfig}
\usepackage{amsmath, amssymb}
\usepackage{verbatim}
\usepackage{multirow}

\begin{document}

\title{Bilayer quantum Hall phase transitions and the 
\\ orbifold non-Abelian fractional quantum Hall states}

\author{Maissam Barkeshli\footnote{Present address: \it Department of Physics, Stanford University, Stanford, CA 94305\rm}}
\author{Xiao-Gang Wen}
\affiliation{Department of Physics, Massachusetts Institute of Technology,
Cambridge, MA 02139, USA }

\begin{abstract}

We study continuous quantum phase transitions that can occur in bilayer 
fractional quantum Hall (FQH) systems as the interlayer tunneling and 
interlayer repulsion are tuned. We introduce a slave-particle gauge theory 
description of a series of continuous transitions from the $(ppq)$ Abelian bilayer states 
to a set of non-Abelian FQH states, which we dub the orbifold FQH states, of which 
the $Z_4$ parafermion (Read-Rezayi) state is a special case. This provides an example 
in which $Z_2$ electron fractionalization leads to non-Abelian topological phases.
The naive ``ideal'' wave functions and ideal Hamiltonians associated with 
these orbifold states do not in general correspond to incompressible phases, 
but instead lie at a nearby critical point. We discuss this unusual situation 
from the perspective of the pattern of zeros/vertex algebra frameworks and
discuss implications for the conceptual foundations of these approaches. 
Due to the proximity in the phase diagram of these non-Abelian states to the 
$(ppq)$ bilayer states, they may be experimentally relevant, both as candidates
for describing the plateaus in single-layer systems at filling fraction 
$8/3$ and $12/5$, and as a way to tune to non-Abelian states in double-layer
or wide quantum wells. 

\end{abstract}

\maketitle

\section{Introduction}

The discovery of topologically ordered phases over the past three decades 
has revolutionized our fundamental understanding of the quantum states of matter.\cite{Wen04}
For a long time, it was believed that different states can be fully classified
by their patterns of symmetry breaking, and transitions between states of different
symmetry can be described through the concept of local order parameters and
the associated Ginzburg-Landau theory of symmetry-breaking. However, the discovery
of the quantum Hall effect showed that even when we break
all symmetries of a system explictly, there can still be distinct quantum 
states of matter that cannot be connected to each other without passing through
a phase transition. These different states are distinguished not by symmetry-breaking order, 
but by a totally different kind of order, called topological order. Understanding
the topological phase of a system at equilibrium is, from one perspective, the coarsest and most basic question
that can be asked of a quantum many-body system, because the result is independent
of any particular symmetry of the problem. In this sense, almost all known conventional 
states of matter -- superfluids, crystals, magnets, insulators, \it etc \rm -- are 
topologically trivial states: if all symmetries of the system are broken, 
most known systems would not have any phase transition as the system parameters are
tuned. 

In order to have a fully developed theory of topological order, we need to understand
how to characterize physically the possible topological states of 
quantum many-body systems, we need a mathematical
framework that describes their properties, and we need to understand how
to describe transitions between different topological states. While we 
know the mathematical framework\footnote{Strictly speaking, this
statement is true for bosonic systems. For fermionic systems, even the
mathematical framework is under development. For a recent discussion,
see \Ref{GW1017}.} -- tensor category theory\cite{W10,P04} -- we do not completely know
how to characterize them physically or how to describe
their phase transitions. This is particularly true in the context of
non-Abelian topological phases. Attempts to develop a systematic 
physical classification of non-Abelian topological orders in FQH states has appeared recently
in the context of the pattern of zeros and vertex algebra approaches to classifying
ideal FQH wave functions.\cite{MR9162,RR9984,WW0808,WW0809,BW0932,BW0937,LW1024} Non-Abelian states are currently the subject of major
experimental and theoretical focus, largely because of the possibility of utilizing
them for robust quantum information storage and processing.\cite{K032,FKL0331,DKL0252}  

This paper presents three main conceptual advances. First, we develop a theory
of a set of continuous quantum phase transitions in bilayer quantum Hall systems between well-known
Abelian states -- the $(ppq)$ states\cite{H8375} -- and a set of non-Abelian topological phases
that we call the orbifold FQH states. This generalizes the discovery in \Ref{BW102}
regarding transitions between the $(p,p,p-3)$ states and the non-Abelian $Z_4$ parafermion 
states. These are all transitions at the same filling fraction
and can be driven by tuning interlayer tunneling and/or interlayer repulsion. These
results are theoretically significant because aside from this series of transitions, there is only
one other set of transitions involving non-Abelian FQH states that is theoretically 
understood; this is the transition between the $(p,p,p-2)$ bilayer states and the Moore-Read
Pfaffian states.\cite{W0050,RG0067} The transitions presented here have experimental consequences: 
we see that there is a possibility of obtaining a wide array of possible non-Abelian
states in bilayer or wide quantum wells by starting with well-known states such
as the $(330)$ state, and tuning the interlayer tunneling and/or interlayer repulsion. 
Furthermore, the non-Abelian states that we present here may also be relevant
in explaining the single-layer plateaus seen in the second Landau level, such as at
$\nu = 8/3$ and $\nu = 12/5$. 

The second major advance relates to the implications of the
orbifold FQH states for the pattern of zeros/vertex algebra
classification.  Currently, the pattern of zeros/vertex
algebra approaches have a shortcoming: some patterns of
zeros (called ``sick'' pattern of zeros) cannot be used to
uniquely fix the ground state and/or quasiparticle wave
functions of FQH states.  Those ``sick'' patterns of zeros
may correspond to gapless states for the ideal Hamiltonian,
leaving open the question of how such solutions may be
relevant in describing gapped, incompressible phases. The
orbifold FQH states that we study here are significant for
the theoretical foundations of the pattern of zeros/vertex
algebra approach because the orbifold states are closely
related to such sick pattern of zeros solutions. While we
use effective field theory and slave-particle gauge theory
techniques to demonstrate the existence of these phases, the
``ideal wave functions'' associated with most of these
orbifold FQH states correspond to the sick pattern of zeros
solutions and are therefore gapless.  In this paper we will
discuss how to properly understand these sick pattern of zeros
solutions and how they are actually relevant to describing
gapped FQH states. 

Finally, the study in this paper shows how non-Abelian states can be obtained
from a theory of $Z_2$ fractionalization, in which the transitions can be viewed
as the condensation of a $Z_2$ charged field while the non-Abelian excitations
correspond to the $Z_2$ vortices. This suggests possible generalizations 
to transitions between Abelian and non-Abelian states based on other discrete
gauge groups. 

The results presented in this paper rely on a diverse, often complementary, 
array of techniques: Chern-Simons (CS) theory, slave-particle methods, conformal
field theory, and vertex algebra. Each technique by itself is not powerful
enough, but the confluence of all them allows us to see the underlying 
structure and to establish our results. 

We will begin in Section \ref{CSTheorySec} by briefly reviewing the
results of an analysis of a particular topological field theory: the
$U(1) \times U(1) \rtimes Z_2$ CS theory, which suggests the possible
existence of a class of non-Abelian FQH states -- the orbifold states. 
However, the topological field theory alone does not imply that there is a
possible FQH state of bosons or fermions with such topological properties.
In Section \ref{slaveSect}, we will develop a slave-particle gauge theory
of $Z_2$ fractionalization that shows that in principle there can be
FQH states whose low energy effective field theories are the 
$U(1) \times U(1) \rtimes Z_2$ CS theories. This slave-particle construction
will yield projected trial wave functions for the orbifold FQH states. 
In Section \ref{edgeSec}, we study the edge theory of these orbifold FQH states
and we develop a prescription for computing all topological quantum numbers
of these phases. We present the results of this prescription in Section \ref{QPcontent}. 

In Section \ref{phaseTransition}, we study the phase transition between the
bilayer Abelian $(ppq)$ states and the orbifold states. We find that the
transition is continuous and in the 3D Ising universality class; the critical
theory is a $Z_2$-gauged Ginzburg-Landau theory. These results give a physical
manifestation of recent mathematical ideas of boson condensation in tensor category
theory.\cite{BS0916}

In Section \ref{vertexAlg} we study the consequences of our results for the pattern of zeros/vertex algebra
approaches to classifying FQH states. Ideal wave functions are wave functions that 
can be obtained through correlation functions of vertex operators in a conformal field
theory; the naive ones for the orbifold FQH states are in general gapless and correspond
to various sick pattern of zeros solutions. We discuss how to interpet this situation
in the vertex algebra framework. The results show how the sick pattern of zeros/vertex
algebra solutions should generally be viewed, and how they are relevant to describing
gapped FQH states even when their associated ideal Hamiltonians are gapless. 

In Section \ref{expSec} we briefly discuss some experimental consequences of this work
and we conclude in Section \ref{conclusion}.

\section{$U(1) \times U(1) \rtimes Z_2$ CS theory and orbifold FQH states }
\label{CSTheorySec}

The $U(1) \times U(1) \rtimes Z_2$ CS theory was introduced in \Ref{BW1023} 
and many of its topological properties were explicitly calculated. Here we 
give a brief review of the main results and pose the main questions that emerge.
The Lagrangian is given by:
\begin{align}
\label{lagrangian}
\mathcal{L} = \frac{p}{4\pi} (a \partial a + \tilde{a} \partial \tilde{a}) + \frac{q}{4\pi} (a \partial \tilde{a} + \tilde{a} \partial a), 
\end{align}
where $a$ and $\tilde{a}$ are two $U(1)$ gauge fields defined in 2+1 dimensions, 
$a \partial a \equiv \epsilon^{\mu \nu \lambda} a_\mu \partial_\nu a_\lambda$, and 
there is an additional $Z_2$ gauge symmetry associated with interchanging
the two gauge fields. The semi-direct product $\rtimes$ highlights the fact that
the $Z_2$ transformation of interchanging the two $U(1)$ gauge fields does not commute with the
individual $U(1) \times U(1)$ gauge transformations. 

In the absence of the $Z_2$ gauge symmetry, (\ref{lagrangian})
is a $U(1) \times U(1)$ CS theory and is the low energy effective field theory for a
bilayer $(ppq)$ FQH state,\cite{BW9045,Wtoprev} where the currents in the two layers are given by:
\begin{align}
j^\mu = \frac{1}{2\pi} \epsilon^{\mu \nu \lambda} \partial_\nu a_\lambda,
\nonumber \\
\tilde{j}^\mu = \frac{1}{2\pi} \epsilon^{\mu \nu \lambda} \partial_\nu \tilde{a}_\lambda.
\end{align}
In the presence of the $Z_2$ gauge symmetry and for $|p - q| > 1$, 
this theory describes a non-Abelian topological phase where $Z_2$ 
vortices are the fundamental non-Abelian excitations. Note that
we use the same Lagrangian for both the $U(1)\times U(1)$ and the 
$U(1) \times U(1) \rtimes Z_2$ CS theories, even though they have 
different gauge structures and are therefore different topological
theories. 

When $p - q = 3$, all of the topological properties of the $U(1) \times U(1) \rtimes Z_2$
CS theory that we can compute agree precisely with those of the $Z_4$ parafermion FQH states at 
filling fraction $\nu = 2/(2q+3)$. This, in conjunction with a number of other results,
led us to suggest that for $p - q = 3$, this is the correct effective field 
theory for the $Z_4$ parafermion FQH states. 

This leads us to ask whether, for more general choices of $p - q$, this theory also describes
a valid, physically realistic, topological phase. In other words,
does it describe a topological phase that can be realized, for some range of material parameters,
in a physical system with realistic interactions? It is not clear because, aside from $p-q = 3$, 
there are no known trial wave functions or trial Hamiltonians that 
capture the properties of such a topological phase. In fact, the naive trial
wave functions that are suggested from a projective construction analysis of these
states are believed to be gapless for $p - q > 3$, which casts doubt 
on whether the phases described by the field theory are physical.
A topological field theory by itself is not enough to know that it 
can be obtained from a physical system of interacting fermions or bosons.
In this paper we will remedy this problem. 

In order to develop the theory for the topological phases that are described
by $U(1) \times U(1) \rtimes Z_2$ CS theory, in the following we will recall
some of the topological properties of such a CS theory. 

\subsection*{Topological properties of $U(1) \times U(1) \rtimes Z_2$ CS theory}

The number of topologically distinct quasiparticles of the $U(1) \times U(1) \rtimes Z_2$ theory
is given by the ground state degeneracy on a torus, which was calculated to be:\cite{BW1023}
\begin{align}
\label{numQP}
\text{No. of quasiparticles } = (N + 7)|p+q|/2,
\end{align}
where  $N \equiv |p-q|$. On genus $g$ surfaces, the number 
of degenerate ground states was calculated to be
\begin{align}
\label{genusgDeg}
S_g(p,q) = |p+q|^g 2^{-1} [ N^g + 1 + (2^{2g}-1)(N^{g-1} + 1)].
\end{align}
Using $S_g(p,q)$, we can read off an important set of topological quantum numbers of the
phase: the quantum dimensions of all the quasiparticles. The quantum
dimension $d_i$ of a quasiparticle of type $i$ has the following meaning. In the presence of $m$
quasiparticles of type $i$ at fixed locations, the dimension of the Hilbert space
grows like $\propto d_i^m$. Abelian quasiparticles have quantum dimension $d = 1$, while
non-Abelian quasiparticles have quantum dimension $d > 1$. 

The ground state degeneracy on genus $g$ surfaces is related to the quantum dimensions through
the formula\cite{V8860,BW0932}
\begin{align}
\label{qdDeg}
S_g = D^{2(g-1)}\sum_{i = 0}^{N_{qp}-1} d_{i}^{-2(g-1)},
\end{align}
where $N_{qp}$ is the number of quasiparticles, $d_i$ is the quantum dimension of the $i$th
quasiparticle, and $D = \sqrt{\sum_i d_i^2}$ is the total quantum dimension. Using 
(\ref{genusgDeg}) and (\ref{qdDeg}), we can calculate the quantum dimension $d_i$ for each
quasiparticle by studying the $g \rightarrow \infty$ limit. The results are as
follows. The total quantum dimension is
\begin{align}
D^2 = 4 N |p+q|.
\end{align}
There are $2|p+q|$ quasiparticles of quantum dimension 1, $2|p+q|$ quasiparticles
of quantum dimension $\sqrt{N}$, and $(N-1)|p+q|/2$ quasiparticles of quantum 
dimension 2.

\begin{figure}[tb]
\centerline{
\includegraphics[scale=0.4]{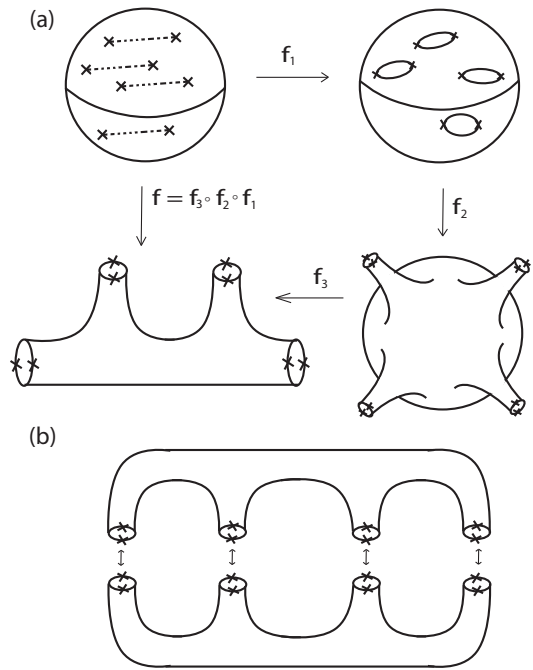}
}
\caption{
\label{Z2vorticesFig}
How to see that $Z_2$ vortices at fixed locations come with
a degeneracy of states and should therefore be non-Abelian.
In (a) we deform the space arbitrarily, and in (b) we
consider a doubled-space with genus $g = n-1$, where $n$ is
the number of pairs of $Z_2$ vortices. Because of the
boundary conditions of the two gauge fields $a$ and
$\tilde{a}$, we can define a single continuous gauge field
on the doubled space, which leaves us with $U(1)$ CS theory
on a genus $g = n-1$ surface. Thus the number of states in
the presence of $n$ pairs of $Z_2$ vortices grows
exponentially in $n$. In this sense, $Z_2$ vortices are like
``genons:'' they effectively change the genus of the
manifold. 
}
\end{figure}

The fundamental non-Abelian excitations in the  $U(1) \times
U(1) \rtimes Z_2$ CS theory are $Z_2$ vortices, which are
topological defects around which the two gauge fields
transform into each other. We can understand the fact that
the $Z_2$ vortices are non-Abelian by seeing that there
should be a degeneracy of states associated with a number of
$Z_2$ vortices at fixed locations. To see that there should
be a degeneracy, see Figure \ref{Z2vorticesFig}. The
configurations of the two gauge fields $a$ and $\tilde{a}$
on a sphere with $Z_2$ vortices can be re-interpreted as
though there is a single gauge field on a doubled space with
genus $g = n-1$, where $n$ is the number of pairs of $Z_2$
vortices. 

In \Ref{BW1023}, we studied the number of degenerate ground
states in the presence of $n$ pairs of $Z_2$ vortices at fixed locations on a sphere. The result 
for the number of such states is:
\begin{equation}
\label{Z2vortexDeg}
\alpha_n =  \left\{
\begin{array}{ll}
    (N^{n-1} + 2^{n-1})/2 &\mbox{ for $N$ even,} \\
    (N^{n-1} + 1)/2 & \mbox{ for $N$ odd.} \\
    \end{array} \right.
\end{equation}
This shows that the quantum dimension of the $Z_2$ vortices is $d = \sqrt{N}$. 
We can also compute the number of states that are odd under the $Z_2$ gauge
transformation. The number of these $Z_2$ non-invariant states also 
turns out to be an important quantity, because it yields important 
information about the fusion rules of the quasiparticles.
The number of $Z_2$ non-invariant states yields the number of ways for $n$ pairs
of $Z_2$ vortices to fuse to an Abelian quasiparticle that
carries $Z_2$ gauge charge. The ground state degeneracy of $Z_2$ non-invariant 
states in the presence of $n$ pairs of $Z_2$ vortices at fixed locations on a sphere
was computed to be
\begin{equation}
\label{Z2noninv}
\beta_n =  \left\{ 
\begin{array}{ll}
    (N^{n-1} - 2^{n-1})/2 &\mbox{ for $N$ even,} \\
    (N^{n-1} - 1)/2 & \mbox{ for $N$ odd.} \\
    \end{array} \right. 
\end{equation}
Thus if $\gamma$ labels a $Z_2$ vortex, these calculations reveal the
following fusion rules for $\gamma$ and its conjugate $\bar \gamma$:
\begin{align}
\label{Z2vortexFusion}
(\gamma \times \bar \gamma)^n = \alpha_n \mathbb{I} + \beta_n j + \cdots, 
\end{align}
where $j$ is a topologically non-trivial excitation that carries the $Z_2$
gauge charge.  The $\cdots$ represent additional quasiparticles that may
appear in the fusion.  

In what follows, we will focus on the case $p - q > 0$, because these are
the cases that are relevant for the bilayer $(ppq)$ states. 

\section{Slave Particle Gauge Theory and $Z_2$ Fractionalization}
\label{slaveSect}

The $U(1) \times U(1) \rtimes Z_2$ CS theory presented above defines a topological
field theory, however for $N \neq 3$ it is unclear whether it can arise as the low 
energy effective field theory of a physical system with local interactions. In this section,
we show how the $U(1) \times U(1) \rtimes Z_2$ CS theory can arise from a slave-particle 
formulation, which adds strong evidence to the possibility of these states being realized in
physical systems with local interactions. The slave-particle formulation provides 
us with candidate many-body wave functions that capture the topological properties of these phases.
It also provides a UV-completion, or lattice regularization, of the
$U(1) \times U(1) \rtimes Z_2$ CS theory. This is useful for computing certain topological
properties, such as the electric charge of the $Z_2$ vortices, which we were unable to calculate
directly from the $U(1) \times U(1) \rtimes Z_2$ CS theory alone. Finally, this
slave-particle formulation provides us with an example in which $Z_2$ electron fractionalization
may lead to non-Abelian topological phases. 

Consider a bilayer quantum Hall system, and suppose that the electrons move on a lattice. 
Let $\Psi_{i\sigma}$ denote the electron annihilation operator at site $i$; 
$\sigma = \uparrow$, $\downarrow$ refers to the two layers. Now consider the positive
and negative combinations:
\begin{align}
\label{Psipm}
\Psi_{i\pm} &= \frac{1}{\sqrt{2}} (\Psi_{i\uparrow} \pm \Psi_{i\downarrow}).
\end{align}

We will use a slave-particle decomposition to rewrite
$\Psi_{\pm}$ in terms of new bosonic and fermionic degrees
of freedom, including appropriate constraints so as not to
unphysically enlarge the Hilbert space. Such slave-particle
decompositions allow us to access novel fractionalized
phases. In the following section, we will introduce a slave
Ising construction that interpolates between the bilayer
Abelian $(ppq)$ states and the states described by the $U(1)
\times U(1) \rtimes Z_2$ CS theory. In Appendix \ref{slaveRotorApp},
we will introduce a slave rotor construction, which can
describe these two phases with the advantage of including a
larger set of fluctuations about the slave-particle
mean-field states. 

\subsection*{Slave Ising}

We introduce two new fields at each lattice site $i$: 
an Ising field $s^z_i = \pm 1$ and a fermionic field $c_{i-}$, 
and we rewrite $\Psi_{i-}$ as
\begin{align}
\Psi_{i+} \equiv c_{i+}, \;\;\;\;
\Psi_{i-} = s^z_i c_{i-}.
\end{align}
This introduces a local $Z_2$ gauge symmetry, associated with the transformations
\begin{align}
s^z_i \rightarrow -s^z_i, \;\;\;\; c_{-_i} \rightarrow -c_{-_i}.
\end{align}
The electron operators are neutral under this $Z_2$ gauge symmetry, and therefore
the physical Hilbert space at each site is the gauge-invariant set of states at
each site:
\begin{align}
(|\uparrow \rangle + |\downarrow \rangle) &\otimes |n_{c_{-}} = 0 \rangle
\nonumber \\
(|\uparrow \rangle - |\downarrow \rangle) &\otimes |n_{c_{-}} = 1 \rangle,
\end{align}
where $|\uparrow \rangle$ ($|\downarrow \rangle$) is the state with
$s^z = +1 (-1)$, respectively. In other words, the physical states at each 
site are those which satisfy
\begin{align}
\label{IsingConstraint}
(s^x_i + 1)/2 + n_{c_{i-}}  = 1.
\end{align}

If we imagine that the fermions $c_{i \pm}$ form some gapped state, then we 
would generally expect two distinct phases:\cite{SF0050} the deconfined/$Z_2$ unbroken
phase, where
\begin{align}
\langle s^z_i \rangle = 0,
\end{align}
and the confined/Higgs phase, where upon fixing a gauge we have
\begin{align}
\langle s^z_i \rangle \neq 0.
\end{align}
We seek a mean-field theory where the deconfined phase corresponds to the 
orbifold FQH states, and the confined/Higgs phase corresponds to the bilayer 
$(ppq)$ states. To do this, observe that in the Higgs phase, we have
\begin{align}
\Psi_{i \pm} = c_{i \pm}, 
\end{align}
since we may set $s^z_i = 1$ in this phase. In such a situation, 
we can use the parton construction\cite{Wnab,W9927} to obtain the $(ppq)$ states.
For example, to obtain the $(330)$ states, we rewrite the electron operators in each layer 
in terms of three partons:
\begin{align}
\Psi_{i\uparrow} &= \psi_{1i} \psi_{2i} \psi_{3i}
\nonumber \\
\Psi_{i\downarrow} &= \psi_{4i} \psi_{5i} \psi_{6i},
\end{align}
where $\psi_a$ carries electric charge $e/3$. We can then rewrite the theory
in terms of the original electrons in terms of a theory of these partons, 
with the added constraint that 
\begin{align}
n_{1i} = n_{2i} = n_{3i}, \;\;\;\;\; n_{4i} = n_{5i} = n_{6i},
\end{align}
where $n_{ai} = \psi^\dagger_{ai} \psi_{ai}$, in order to preserve 
the electron anti-commutation relations and to avoid
unphysically enlarging the Hilbert space at each site. The $(330)$ state corresponds
to the case where each parton forms a $\nu = 1$ integer quantum Hall
state. 

Therefore, to interpolate between the $Z_4$ parafermion state and the
$(330)$ state at $\nu = 2/3$, we write:
\begin{align}
\Psi_{i+} &=  \psi_{1i} \psi_{2i} \psi_{3i} + \psi_{4i} \psi_{5i} \psi_{6i}, 
\nonumber \\
\Psi_{i-} &= s^z_i ( \psi_{1i} \psi_{2i} \psi_{3i} - \psi_{4i} \psi_{5i} \psi_{6i}).
\end{align}
More generally, to describe the $(ppq)$ states and the orbifold FQH states,
we set
\begin{align}
\label{slaveIsing}
\Psi_{i+} &= c_{i+}, \;\;\;\;
\Psi_{i-} = s^z_i c_{i-}
\nonumber \\
c_{i\pm} &= \left( \prod_{a = 1}^N \psi_{ai} \pm \prod_{a = N+1}^{2N} \psi_{ai} \right) 
\prod_{b = 2N+1}^{2N+q} \psi_{bi},
\end{align}
where $N \equiv p-q$ (note that we assume $p > q$). Furthermore, we 
assume that the interactions are such that the partons each form a 
$\nu = 1$ IQH state. 

\subsubsection*{Topological properties of the $Z_2$ confined and deconfined phases}
\label{topProperties}

In what follows, let us focus on the case $q = 0$. 
When $\langle s^z_i \rangle = 1$, we can write 
\begin{align}
\Psi_{i\pm} = \prod_{a = 1}^N \psi_{ai} \pm \prod_{a = N+1}^{2N} \psi_{ai} .
\end{align}
The low energy theory will thus be a theory of $2N$ partons, each with electric
charge $e/N$, and coupled to a $SU(N) \times SU(N)$ gauge field:
\begin{align}
\label{partonL}
\mathcal{L} = i \psi^{\dagger} \partial_0 \psi
+ \frac{1}{2m} \psi^{\dagger} (\partial - i A_i Q)^2 \psi 
+ \text{Tr }(j^\mu a_\mu) + \cdots,
\end{align}
where $a$ is an $SU(N) \times SU(N)$ gauge field,
$\psi^{\dagger} = (\psi_{1}^{\dagger}, \cdots, \psi_{2N}^{\dagger})$,
$(Q)_{ab} = \delta_{ab} e/N$, $A$ is the external electromagnetic
gauge field, and $j^\mu_{ab} = \psi^{\dagger}_a \partial^\mu \psi_b$. 
If the partons form a $\nu = 1$ IQH state, then we can integrate 
out the partons to obtain a $SU(N)_1 \times SU(N)_1$ CS theory as 
the long-wavelength, low energy field theory. This $SU(N)_1 \times SU(N)_1$ 
CS theory reproduces all of the correct ground state properties, such as 
the ground state degeneracy on genus $g$ surfaces, and the fusion rules of the
quasiparticles. The quasiparticle excitations are related to holes in the parton
integer quantum Hall states. The $SU(N)_1 \times SU(N)_1$ CS theory needs
to be supplemented with additional information about the fermionic character of
an odd number of holes in order to completely capture all of the topological
quantum numbers. This can be done by using the $U(1)_N \times U(1)_N$ CS theory
instead, which is known to be the correct low-energy effective field theory
of the bilayer $(NN0)$ states. 

Now consider the $Z_2$ deconfined phase, where $\langle s^z_i \rangle = 0$; 
what is the low energy effective theory? Since the partons still 
each form a $\nu = 1$ IQH state and are coupled to an $SU(N) \times SU(N)$ gauge field, 
integrating them out  will yield a $SU(N)_1 \times SU(N)_1$ CS theory, and using 
the arguments outlined above, we are left with a $U(1)_N \times U(1)_N$ CS theory. 
Suppose that we also sum over the Ising spins $\{s^z_i\}$. Since there are no 
gapless modes associated with phases of the Ising spins, we expect a local action 
involving the $Z_2$ gauge field coupled to the $U(1)$ gauge fields. We do not know how
to explicitly write this action down, because the CS terms are difficult to properly
define on a lattice, while the discrete gauge fields require a lattice for their action.
Nevertheless, we consider the theory on general grounds: observe that the
$Z_2$ gauge symmetry interchanges $\psi_a$ and $\psi_{a+N}$; thus in the low energy
theory involving only the gauge fields, the $Z_2$ gauge symmetry interchanges 
the current densities associated with the two $U(1)$ gauge
fields. This is precisely the content of the $U(1) \times U(1) \rtimes Z_2$ 
CS theory. Thus, we may think of the $Z_2$ deconfined phase
of this slave Ising construction as providing a UV-completion of the
$U(1) \times U(1) \rtimes Z_2$ CS theory. In a sense, we can even think of this slave particle gauge
theory as the complete definition of the $U(1) \times U(1) \rtimes Z_2$ CS 
theory.\footnote{An alternative, mathematical definition of CS theory 
for disconnected gauge groups is given in \Ref{DW9093}.}

Now let us further study the low energy excitations of this $Z_2$ fractionalized phase. In this 
phase, the Ising spin $s^z$ can propagate freely and is deconfined
from the partons. This is an electrically neutral excitation that is charged under the
$Z_2$ gauge symmetry and that fuses with itself to the identity. 
The phases described by the $U(1) \times U(1) \rtimes Z_2$ CS theory all have precisely
such a $Z_2$ charged excitation; (\ref{Z2noninv}) and (\ref{Z2vortexFusion}) 
yield the number of ways for $n$ pairs of $Z_2$ vortices to fuse to precisely this 
$Z_2$ charged excitation, which was denoted $j$. 

The other novel topologically non-trivial excitation in the $Z_2$ deconfined phase
is the $Z_2$ vortex. Since the $Z_2$ gauge field is coupled to the partons, the 
$Z_2$ vortex is non-Abelian. This is not an obvious result: in the low-energy 
$U(1) \times U(1) \rtimes Z_2$ CS theory the $Z_2$ vortex 
corresponds to a topological defect around which the
two $U(1)$ gauge fields transform into each other. A detailed study of the $Z_2$
vortices in the $U(1) \times U(1) \rtimes Z_2$ theory shows that there is a topological
degeneracy associated with the presence of $n$ pairs of $Z_2$ vortices at fixed locations,
which reveals that the $Z_2$ vortices are non-Abelian quasiparticles (see Figure \ref{Z2vorticesFig}). 

\subsubsection*{Electric charge of $Z_2$ vortices}

Can we understand the allowed values of the electric charge carried by the $Z_2$
vortices? We believe the $U(1) \times U(1) \rtimes Z_2$ CS theory, for certain
choice of coupling constants, describes the $Z_4$ parafermion state. 
The $Z_4$ parafermion state has a fundamental non-Abelian
excitation that carries a fractionalized electric charge; at $\nu = 2/3$ for example,
the electric charge of the fundamental non-Abelian excitation comes in odd multiples of $e/6$.
Since we believe that the $Z_2$ vortices in this theory correspond to the
fundamental non-Abelian excitations, an important check on this 
slave Ising description will be whether it can account
for these values of the fractionalized electric charge. 

To calculate the electric charge, let us define the following parton operators, 
which are superpositions of the parton operators $\psi_a$:
\begin{align}
\psi_{a\pm} = \frac{1}{\sqrt{2}}(\psi_a \pm \psi_{a+N} ), \;\;\;\; a = 1, \cdots, N.
\end{align}
The local $Z_2$ gauge symmetry corresponds to the transformation:
\begin{align}
s^z_i &\rightarrow -s^z_i,
\nonumber \\
\psi_a &\leftrightarrow \psi_{a+N}, \;\;\; a = 1,..,N.
\end{align}
Thus, $\psi_{a+}$ is $Z_2$-neutral while $\psi_{a-}$ is $Z_2$-charged. Furthermore,
since $\psi_a$ each form a $\nu = 1$ IQH state, then $\psi_{a+}$ and $\psi_{a-}$ 
also each form $\nu = 1$ IQH states. The particle/hole excitations of the states formed by
$\psi_{a-}$ carry electric charge $e/N$ (recall we set $q = 0$ in (\ref{slaveIsing})). 
The $Z_2$ vortex acts as a $\pi$-flux for $\psi_{a-}$. Thus in the 
low-energy field theory, the interaction between
the excitations of the $\psi_{a-}$ IQH state and the external electromagnetic gauge
field $A_\mu$ and the $Z_2$ vortices is described by
\begin{align}
\mathcal{L}_{int;-} = \sum_{a=1}^N (\frac{e}{N}A_\mu + b_\mu) j^\mu_{a-},
\end{align}
where a $Z_2$ vortex is associated with $\pi$ flux of the $U(1)$ gauge field $b_\mu$. 
$j^\mu_{a-}$ is the current density associated with the $\psi_{a-}$ partons. Integrating
out the partons, which are in a $\nu = 1$ IQH state, will generate a Chern-Simons term:
\begin{align}
\mathcal{L}_{int;-} &= \sum_{a=1}^N \frac{1}{4\pi}(\frac{e}{N} A + b) \partial (\frac{e}{N} A + b),
\nonumber \\
&= \frac{1}{4\pi} \frac{e^2}{N} A \partial A + \frac{N}{4\pi} b \partial b + \frac{e}{2\pi} A \partial b.
\end{align}
Notice that the interaction between the $\psi_{a+}$ current and the external electromagnetic
gauge field will contribute another term $\frac{1}{4\pi} \frac{e^2}{N} A \partial A$ to the
action, from which we see that the filling fraction is $\nu = 2/N$. 
Furthermore, because of the coupling of $b$ to the external gauge field $A$, we see that 
a $\pi$ flux of the $b_\mu$ gauge field will carry charge $e/2$. Therefore, depending on how many
holes, $m$, of the parton integer quantum Hall states are attached to the $Z_2$ vortices, 
the $Z_2$ vortices can have electric charge of 
\begin{align}
Q_{Z_2 \text{ vortex}} = e (2m +N)/2N.
\end{align}
When $N = 3$, this result agrees exactly with properties of the $Z_4$ parafermion
state, which is that the electric charge of the fundamental non-Abelian quasiparticles
comes in odd multiples of $e/6$. More generally, when $N$ is odd (even), 
we see that the $Z_2$ vortices can only carry electric charge in odd (even) 
integer multiples of $e/2N$. In Section \ref{genQPprop}, we will again 
see precisely these results through a totally different description of this phase!

\subsection*{Slave Ising projected wave functions} 

The slave-particle approach naturally suggests trial wave functions that capture the 
essential long-wavelength properties of the phase. First we have the mean-field
state of the partons and the Ising spins:
\begin{align}
|\Phi_{mf} \rangle = |\{s^z_i\}\rangle |\{\psi_a\} \rangle,
\end{align}
where the partons $\psi_a$ form a $\nu = 1$ IQH state. The $Z_2$ 
confined/Higgs phase, which describes the Abelian $(ppq)$ states,
will be associated with an ordered state of the Ising spins. 
The $Z_2$ deconfined phase will be described by an unordered, paramagnetic state of
the Ising spins. The quantum state of the electrons will be given by a projection
onto the physical Hilbert space:
\begin{align}
|\Psi \rangle = \mathcal{P} |\Phi_{mf} \rangle,
\end{align}
where 
\begin{align}
\mathcal{P} &= \prod_i \mathcal{P}_i,
\nonumber \\
\mathcal{P}_i &= \mathcal{P}_i^{Ising} \mathcal{P}_i^{Parton}.
\end{align}
The projection operator for the Ising sector is (see eqn. \ref{IsingConstraint}):
\begin{align}
\mathcal{P}_i^{Ising} &= \frac{1}{2}[1 - (-1)^{((s^x_i + 1)/2 + n_{c_{i-}})}],
\end{align}
where $n_{c_{i-}} = c_{i-}^\dagger c_{i-}$ is written in terms of the partons as
\begin{align}
n_{c_{i-}} = & \frac{1}{2}(n_{\uparrow i} + n_{\downarrow i}) - 
\nonumber \\
& \frac{1}{2}[(\psi_{1i} \cdots \psi_{Ni})^{\dagger}(\psi_{N+1 i} \cdots \psi_{2N i}) + h.c.].
\end{align}
$n_{i \uparrow}$ and $n_{i \downarrow}$ are the number of electrons in the top and bottom 
layer, respectively, at site $i$. 
The projection operator for the parton sector is:
\begin{align}
\mathcal{P}_i^{Parton} &= \prod_{a=1}^N [1 - (n_{i\uparrow} - n_{ai})^2] \prod_{a=N+1}^{2N} [1 - (n_{i\downarrow} - n_{ai})^2],
\end{align}
which implements the constraint $n_{1i} = \cdots = n_{Ni} = n_{i\uparrow}$ and 
$n_{N+1i} = \cdots = n_{2Ni} = n_{i\downarrow}$.

Alternatively, we can work with the spatial wave function.
The amplitude of the electron wave function to have 
$N_{\uparrow}$ electrons in one layer and $N_{\downarrow}$ electrons in the
second layer is given by
\begin{align}
\Psi(\{ \v r_i \}, \{ \v r_i' \}) = 
\langle 0 | \prod_{i=1}^{N_{\uparrow}} \Psi_{\uparrow \v r_i} \prod_{i=1}^{N_{\downarrow}} 
\Psi_{\downarrow \v r_i'}|\Phi_{mf} \rangle,
\end{align}
where $\Psi_{\sigma \v r}$ is given in terms of the partons and the Ising spins 
through (\ref{Psipm}) and (\ref{slaveIsing}). Here, 
$|0\rangle = |0\rangle_{parton} |\{s^x_i = 1\}\rangle$ is the state with no partons
and an eigenstate of $\hat{s}^x_i$ with eigenvalue $1$. 

This wave function is important because currently it is the only wave function
we have for these non-Abelian FQH states (for $N > 3$). As we will discuss later, there is currently
no corresponding ideal wave function for these states. The projected wave functions 
presented here can in principle be used for numerical studies in order to determine 
which phases are most likely under realistic physical conditions.

\section{Edge theory of the orbifold FQH states}
\label{edgeSec}

One use of the $U(1) \times U(1) \rtimes Z_2$ CS theory is that it can be used
to study the edge theory of the associated topological phases. It is known that
the $U(1)$ CS description of the Abelian quantum Hall liquids leads to the
chiral Luttinger liquid edge theory.\cite{W9211} More specifically, an $n$-component Abelian quantum
Hall liquid can be described by a CS theory involving $n$ U(1) gauge 
fields:\cite{WZ9290}
\begin{align}
\mathcal{L} = \frac{1}{4\pi} K_{IJ} a_I \partial a_J + \frac{1}{2\pi}A \partial a_I ,
\end{align}
where $K$ is an $n \times n$ symmetric invertible matrix and 
$A$ is the external electromagnetic gauge field. As a result the 
edge theory is described by $n$ chiral free bosons:\cite{W9211}
\begin{align}
\mathcal{L}_{edge} = 
K_{IJ} \partial_t \phi_I \partial_x \phi_J - V_{IJ} \partial_x \phi_I \partial_x \phi_J,
\end{align}
where $V_{IJ}$ is a positive definite matrix that dictates the velocity of the
edge modes and depends on microscopic properties of the edge. 

We therefore expect that the edge of the phases described by 
$U(1) \times U(1) \rtimes Z_2$ CS theory will be described by
two free chiral bosons, $\varphi_1$ and $\varphi_2$, with the
Lagrangian given above, and with an additional $Z_2$ gauge 
symmetry associated with the transformations
\begin{align}
(\varphi_1(z), \varphi_2(z)) \sim (\varphi_2(z), \varphi_1(z))
\end{align}
at each spacetime point. Such a CFT is called an orbifold CFT, because
the symmetry $U(1) \times U(1)$ of the original free boson theory
is gauged by a discrete $Z_2$ symmetry. Thus we refer to this theory 
as the $[U(1) \times U(1)]/Z_2$ orbifold CFT. That 
the $U(1) \times U(1) \rtimes Z_2$ CS theory should correspond 
to this edge CFT may be expected in light of Witten's CS/CFT 
correspondence.\cite{W8951,MS8922,DW9093} 

As a check, we may perform a simple counting of the operator content
of such a chiral CFT by following the considerations of \Ref{DV8985}. In that
reference, it was argued that the number of primary operators (primary with
respect to the orbifold chiral algebra) in a $G/Z_k$ orbifold CFT is related
to the number of primary operators in the un-orbifolded CFT, with symmetry group $G$, 
by the formula
\begin{align}
\text{No. of operators } = nk^2 + m.
\end{align}
Here, $m$ is the number of groups of $k$ operators in the original un-orbifolded theory
that are cyclically permuted by the $Z_k$ action; together, they lead to $m$ operators 
that are $Z_k$ invariant. $n$ is the number of operators in the original un-orbifolded
theory that are fixed under the $Z_k$ action. 

In the case of the orbifold states with $p - q = N$ and $q = 0$, the
primary operators are labelled as 
\begin{align}
V_{ab}(z) = e^{i a/\sqrt{N} \varphi_1(z) + i b /\sqrt{N} \varphi_2(z)}.
\end{align}
The $Z_2$ action exchanges $a$ and $b$, so we have $n = N$ and $m = N(N-1)/2$. 
This leads to $N(N+7)/2$ primary operators, which agrees exactly with the number of 
quasiparticles expected from the analysis of the torus ground state 
degeneracy of the $U(1) \times U(1) \rtimes Z_2$ CS theory. Carrying out 
the calculation for general $q \neq 0$ yields
\begin{align}
\text{No. of operators } = (N+7)|p+q|/2,
\end{align}
again agreeing with the analysis from the $U(1) \times U(1) \rtimes Z_2$
CS theory (see eqn. (\ref{numQP})). This highly non-trivial consistency 
check suggests that this is indeed the correct edge theory. 

In order to obtain the full topological properties of these FQH states using
the edge theory, we would need to obtain the scaling dimensions of each
of the primary operators and their fusion rules. This can be done by first
computing the characters of the chiral CFT, which are given by
\begin{align}
\chi_i(\tau) = Tr_{[\mathcal{O}_i]} q^{L_0 - c/24}.
\end{align}
The trace is over states in the module labelled $[\mathcal{O}_i]$, where 
$\mathcal{O}_i$ is a primary field of the chiral algebra. For FQH states,
the $\mathcal{O}_i$ label different quasiparticle sectors. $L_0$ is the generator
of scale transformations, $q = e^{2\pi i \tau}$, and $c$ is the central  charge. 
The scaling dimensions and fusion rules of the primary fields can be obtained by
studying the transformation rules of the characters under the modular transformations
$S: \tau \rightarrow -1/\tau$ and $T: \tau \rightarrow \tau + 1$.\cite{FMCFT}

Using the chiral characters, we would also be able to obtain the
full edge spectrum for the FQH states on a disk. The spectrum of edge states
at each angular momentum in the topological sector labelled by $\mathcal{O}_i$
are given by the coefficients $a_n^{(i)}$ in the expansion
\begin{align}
\chi_i(\tau) = \sum_{n = 0}^\infty a_n^{(i)} q^n.
\end{align}
Here, $a_n^{(i)}$ is the number of edge excitations with
energy $E_n\propto n$ on a disk with the quasiparticle
created by $\mathcal{O}_i$ at the center of the disk.

Unfortunately, obtaining the characters is a highly
non-trivial task.  In known examples of orbifold CFTs, one
common way to proceed is to first compute the torus
partition function of the non-chiral theory, which includes
both holomorphic and anti-holomorphic sectors of the CFT.
The partition function is related to the chiral characters
through
\begin{align}
Z(\tau, \bar \tau) = \sum_{ij} \chi_i(q) M_{ij} \bar{\chi}_j(\bar q),
\end{align}
where $M$ is a matrix that specifies how to glue together the holomorphic and
anti-holomorphic sectors. Sometimes it is possible to take the ``holomorphic square root'' and guess the
chiral characters $\chi_i(\tau)$ from the partition function $Z(\tau, \bar \tau)$. 
This is done, for example, for the $U(1)/Z_2$ orbifold CFTs at $c = 1$.\cite{DV8985,FMCFT}
In the case of FQH states, $Z(\tau, \bar \tau)$ may be used to compute the edge
spectrum on a cylinder by expanding in powers of $q$ and $\bar q$. 

In the case of the $[U(1) \times U(1)]/Z_2$ orbifold CFT, it is possible
to compute $Z(\tau, \bar \tau)$, but we do not know at present how to take the
holomorphic square root and thus derive the scaling dimensions and fusion rules
of the operators in the edge theory. In spite
of this shortcoming, we can develop a prescription for computing the 
scaling dimensions and fusion rules of the operators in this 
CFT. We will perform many highly non-trivial checks with both the slave particle
gauge theory and with results of the $U(1) \times U(1) \rtimes Z_2$ CS theory
in order to confirm that the prescription given yields correct results. This
prescription is necessary because it is currently the only way we have of computing
all of the topological quantum numbers of the orbifold states. While the slave
Ising and associated $U(1) \times U(1) \rtimes Z_2$ CS theory descriptions are
powerful and can be used to calculate many highly non-trivial topological properties,
we do not currently know how to use them to compute all topological
properties of the orbifold states, such as the spin of the $Z_2$ vortices or the
full set of fusion rules. 



First observe that if we consider the following combination of the chiral scalar fields:
\begin{align}
\varphi_{\pm} = \frac{1}{\sqrt{2}} (\varphi_1 \pm \varphi_2),
\end{align}
Then the action becomes equivalent to the action of a free chiral scalar
field, $\varphi_+$, and that of the $U(1)/Z_2$ orbifold, described by $\varphi_{-}$. 
However, the edge theory is not simply a direct product of these two independent
theories. The reason is that the fields $\varphi_1$ and $\varphi_2$ are compactified:
in the case $q = 0$, we have
\begin{align}
\varphi_i \sim \varphi_i + 2\pi R.
\end{align}
The compactification radius $R$ is related to $N$ through $R^2 = N$. 

The spectrum of compactified bosons includes winding sectors; on a torus with
spatial length $L$, the bosons can wind:
\begin{align}
\varphi_i(x+L,t) = \varphi_i(x,t) + 2\pi R.
\end{align}
As a result, the fields $\varphi_+$ and $\varphi_-$ are not independent and instead are
tied together by their boundary conditions. We may think of such a theory, which is
equivalent to the $[U(1) \times U(1)]/Z_2$ as a theory denoted by
$U(1) \otimes U(1)/Z_2$. The $\otimes$ indicates the non-trival gluing together of
the $U(1)$ theory and the $U(1)/Z_2$ orbifold theory. Let us consider the gluing
together of these two theories from the point of view of the chiral operator algebra. 

Observe that the edge theory for the $(NN0)$ states is generated by the electron
operators
\begin{align}
\Psi_{e1}(z) = e^{i \sqrt{N} \phi_1(z)}, \;\;\;\; \Psi_{e2}(z) = e^{i \sqrt{N} \phi_2(z)},
\end{align}
where $\phi_1$ and $\phi_2$ are free scalar bosons in a 1+1D chiral CFT. 
In terms of $\varphi_{\pm}$, we have
\begin{align}
\Psi_{e1} = e^{i \sqrt{N/2} ( \varphi_+ + \varphi_- )}, \;\;\;\; \Psi_{e2} = e^{i \sqrt{N/2} (\varphi_+ - \varphi_-)}.
\end{align}
$\varphi_+$ describes the electrically charged sector of the edge theory, 
while $\varphi_-$ describes the neutral sector of the edge theory. More generally,
for the $(ppq)$ states, the electron operators in the top and bottom layers are:
\begin{align}
\Psi_{e1} &= e^{i \sqrt{N/2} \varphi_-} e^{i \sqrt{\frac{p+q}{2}}  \varphi_+ }, 
\nonumber \\
\Psi_{e2} &= e^{-i \sqrt{N/2} \varphi_-} e^{i \sqrt{\frac{p+q}{2}} \varphi_+},
\end{align}
where recall $N = p - q > 0$.

The chiral algebra of the $[U(1) \times U(1)]/Z_2$ theory should be the
$Z_2$ invariant subalgebra of the $U(1) \times U(1)$ chiral algebra. 
Therefore we expect it to be generated by
\begin{align}
\Psi_{e+} \propto \Psi_{e1} + \Psi_{e2} \propto \cos(\sqrt{N/2} \varphi_-) e^{i \sqrt{\frac{p+q}{2}} \phi_+}.
\end{align}
Studying the chiral algebra of $\Psi_{e+}$ should yield the spectrum of edge states; 
representations of this chiral algebra should yield the topologically distinct sectors in the
edge theory and should correspond to the topologically inequivalent quasiparticles in the
bulk. The OPE $\Psi_{e+}^\dagger(z) \Psi_{e+}(w)$ contains only operators even in $\varphi_-$. 
In particular, it contains the operator $\cos(\sqrt{2N} \varphi)$, which is known to generate the 
chiral algebra of the $U(1)_{2N}/Z_2$ orbifold CFT.\cite{DV8985} Note that the level $2N$ is related
to the compactification radius of the boson -- see Appendix \ref{orbAppendix} for a review. 
The chiral algebra of this orbifold CFT is denoted 
$\mathcal{A}_{N}/Z_2$, where $\mathcal{A}_N$ is the chiral algebra of the $U(1)_{2N}$
Gaussian theory. $\mathcal{A}_{N}$ is generated by the operators $\{e^{\pm i \sqrt{2N} \varphi} \}$,
and $\mathcal{A}_{N}/Z_2$ is the $Z_2$ invariant subalgebra of $\mathcal{A}_{N}$, which is 
generated by $\cos(\sqrt{2N} \varphi)$. Focusing on the neutral sector of these FQH edge
theories, we see that the electron operators at the edge of the $(ppq)$ states can generate 
the algebra $\mathcal{A}_N$, while the operator $\Psi_{e+}$ can only 
generate the algebra $\mathcal{A}_N/Z_2$. 

The operator $\cos(\sqrt{N/2} \varphi)$ is difficult to work with for our
purposes, but it is very closely related to the primary field $\phi^1_N$
in the $U(1)/Z_2$ orbifold CFT (see Appendix \ref{orbAppendix} for a detailed
discussion of the operator content in the $U(1)/Z_2$ CFT), which motivates us to use the 
following operator as the electron operator:
\begin{align}
\label{elOp}
\Psi_e(z) = \phi_N^1(z) e^{i \sqrt{(p+q)/2} \phi_c(z)}.
\end{align}
This describes a FQH state at filling fraction $\nu = 2/(p+q)$. 
$\phi_N^1$ is a primary field of the $Z_2$ orbifold chiral algebra with scaling
dimension $N/4$ and its fusion
rules with other primary fields is known, so it is more convenient to work with
$\phi_N^1$ than with $\cos(\sqrt{N/2} \varphi)$. We expect that
both operators could in principle be used to generate the same 
edge spectrum. The chiral algebra of the electron operator will be referred to
as $\mathcal{A}_e$; note that it contains the entire orbifold chiral algebra as a
subalgebra: $\mathcal{A}_N/Z_2 \subset \mathcal{A}_e$. 

Now we make the following conjecture for the edge theory. The properties
of the chiral operators in the $[U(1) \times U(1)]/Z_2$ theory
can be obtained by studying operators in the $U(1)/Z_2 \times U(1)$ 
CFT that are local -- i.e. have a single-valued OPE -- with respect to 
the electron operator (\ref{elOp}). Two operators are topologically equivalent 
if they can be related by an operator in the electron chiral algebra. 
Practically, this means that the topologically distinct quasiparticle operators
$V_{\gamma}$ are of the form 
\begin{align}
V_{\gamma} = \mathcal{O}_\gamma e^{i Q_\gamma \sqrt{\nu^{-1}} \phi_c},
\end{align}
where $\mathcal{O}_\gamma$ is a chiral primary operator from the $U(1)_{2N}/Z_2$
orbifold CFT and determines the non-Abelian properties of the quasiparticle,
and $Q_\gamma$ determines the electric charge of the quasiparticle. 

The quasiparticle operators in the edge theory yield all the topological
properties of the bulk excitations. The scaling dimensions 
$h_\gamma = h_{\mathcal{O}_\gamma} + Q_\gamma^2/2\nu$
of the quasiparticle operators in the CFT are related to an important
topological quantum number of the bulk excitations: the
quasiparticle twist, $\theta_\gamma = e^{2 \pi i h_\gamma}$, which specifies the phase accumulated 
as a quasiparticle is rotated by $2\pi$. The fusion rules 
of the quasiparticles in the bulk are identical to the fusion rules of 
the quasiparticle operators in the edge theory. 

To summarize, the conjecture is that the properties of the chiral primary fields
in the $[U(1) \times U(1)]/Z_2$ CFT can be obtained by instead
considering the electron operator (\ref{elOp}) and embedding the electron 
chiral algebra $\mathcal{A}_e$ into the chiral algebra of the $U(1)_{2N}/Z_2 \times U(1)$
CFT. This allows us to study representations of $\mathcal{A}_e$ in terms of
primary fields in the $U(1)_{2N}/Z_2 \times U(1)$ CFT. 

\section{Quasiparticle content and topological quantum numbers 
of orbifold FQH states}
\label{QPcontent}

Using the above prescription for finding the topologically inequivalent 
quasiparticle operators in CFT, we obtain the complete topological quantum 
numbers that such an edge theory describes. 

Remarkably, the topological properties obtained through this CFT prescription
agree exactly with all the properties that we can compute 
from the $U(1) \times U(1) \rtimes Z_2$ CS theory and the slave Ising theory 
through completely different methods. Below, we will first illustrate 
a simple way of understanding the results obtained from this edge theory
in terms of the $U(1) \times U(1) \rtimes Z_2$ CS theory and in terms 
of the quasiparticle content of the $(ppq)$ states. We will 
then proceed to study some specific examples in more detail.

\subsection*{General properties}
\label{genQPprop}

To illustrate the main ideas, we set $q = 0$. When $q = 0$,
the orbifold FQH states have $N(N+7)/2$ topologically inequivalent
quasiparticles (see eqn.(\ref{numQP})). $2N$ quasiparticles have quantum dimension $1$,
$2N$ have quantum dimension $\sqrt{N}$, and $N(N-1)/2$ have 
quantum dimension 2. 

Label the $d = 1$ and $d = \sqrt{N}$ quasiparticles as $A_l$ and $B_l$,
respectively, for $l = 0, \cdots, 2N-1$. Let us label the $N(N-1)/2$ 
quasiparticles with $d = 2$ as $C_{mn}$, where $m,n = 0, ..., N-1$ and
$m > n$. These quasiparticles have the properties listed in Table \ref{qpContentGen}.
\begin{table}
\begin{tabular}{cccc}
 & Charge & Scaling Dim. & Quantum Dim. \\
\hline
$A_l$  & $2l/N$& \;\;\;\;\; $l^2/N$ \;\;\;\;\; & 1 \\
&&&\\
\multirow{2}{*}{$B_l$}  &$l/N$, $N$ even  &   & \multirow{2}{*}{$\sqrt{N}$}  \\
  &$(2l+1)/2N$, $N$ odd &  & \\
&&&\\
$C_{mn}$ & $(m+n)/N$  & $(m^2 + n^2)/2N$ \;\;\;\;\; & 2 \\
\hline
\end{tabular}
\caption{
\label{qpContentGen}
General properties of quasiparticles in the orbifold FQH states
for $q = 0$. The quasiparticles are labelled here as $A_l$ and $B_l$ for 
$l = 0, .., 2N-1$, and $C_{mn}$ for $m,n = 0, .., N-1$ and $m > n$. 
The $A_l$ and $C_{mn}$ quasiparticles are closely related to the 
Abelian quasiparticles of the $(NN0)$ states, while the $B_l$ quasiparticles
are the $Z_2$ vortices in the $U(1) \times U(1) \rtimes Z_2$ CS theory. 
}
\end{table}
We find that when $N$ is even, the non-Abelian quasiparticles 
$B_l$ have charge $l/N$, and when $N$ is odd, the non-Abelian 
quasiparticles $B_l$ have charge $(2l+1)/2N$.

Now consider the bilayer $(NN0)$ states, which have $N^2$ Abelian quasiparticles
that can be labelled by two integers $(m,n)$, and where 
$(m,n) \sim (m+N,n) \sim (m,n+N)$ all refer to topologically equivalent quasiparticles. 
The electric charge of these quasiparticles is given by $(m+n)/N$ and 
the scaling dimension is given by $(m^2 + n^2)/2N$. 

The quasiparticle content of the orbifold FQH states can now be interpreted
in the following way. $A_l$ for $l = 0, .., N-1$ is the same as the 
quasiparticles $(l,l)$ from the $(NN0)$ states: they are all Abelian, and 
$A_l$ carries the same charge and statistics as $(l,l)$. Furthermore, the orbifold
FQH states have an additional neutral, Abelian boson that squares to the identity.
In terms of the $U(1) \times U(1) \rtimes Z_2$ CS theory, it can be interpreted
as the quasiparticle that carries $Z_2$ gauge charge. The $Z_2$ charged quasiparticle
can fuse with the $A_l$ for $l = 0, .., N-1$ to yield the $A_l$ for $l = N, ..., 2N-1$.
The quasiparticles $C_{mn}$ correspond to the $Z_2$ invariant combinations of
$(m,n)$: $C_{mn} \sim (m,n) + (n,m)$, for $m \neq n$. This is clear 
in the edge theory, where these quasiparticle operators take 
the form $\cos(l \varphi_-/\sqrt{2N}) e^{iQ \sqrt{\nu^{-1}} \phi_c}$.

Finally, the quasiparticles $B_l$ correspond to $Z_2$ vortices in the 
$U(1) \times U(1) \rtimes Z_2$ CS description. Alternatively, in the edge
orbifold theory, they correspond to twist operators.  There is a 
fundamental $Z_2$ vortex, say $B_0$, and the other $B_l$ can be obtained by fusing with 
the $A$ or $C$ quasiparticles. Note that when $N$ is odd, the
minimal quasiparticle charge in the orbifold states is carried
by a $Z_2$ vortex and is given by $1/2N$. This is half of the minimal
quasiparticle charge in the corresponding $(NN0)$ states. 

\subsection*{Examples}

One of the simplest examples of the above properties is shown in 
Table \ref{N3qp}, which describes the quasiparticle content for 
$(N,q) = (3,0)$. When $N = 3$, the orbifold FQH states are the same
as the $Z_4$ parafermion FQH states at filling fraction $\nu = 2/(2q+3)$.
In this example, we clearly see three different
families of quasiparticles, and each family forms a representation of a
magnetic translation algebra.\cite{BW0932} Notice that the quasiparticle 
$j \sim \partial \varphi$ is odd under the $\varphi \sim -\varphi$ 
transformation of the orbifold CFT, which is one way of seeing that 
this quasiparticle should carry $Z_2$ gauge charge in the 
$U(1) \times U(1) \rtimes Z_2$ CS description. 

In Tables \ref{N2qp} and \ref{N4qp} we list the quasiparticle content for
the cases $(N,q) = (2,0)$ and $(4,0)$. These states are slightly more complicated
than the $N = 3$ case because there are more than three representations 
of a magnetic translation algebra and there is not a one-to-one correspondence 
between the pattern of zeros sequences $\{n_l\}$ and topologically inequivalent quasiparticles.
We study these features further in Section \ref{vertexAlg}. 

In Tables \ref{N2qp}-\ref{N4qp}, we have also listed the occupation 
sequences $\{n_l\}$ of each quasiparticle, which are defined as follows.
If $\Psi_e$ is the electron operator and $V_\gamma$ is a quasiparticle operator,
we obtain a sequence of integers $\{l_{\gamma;a}\}$ from the following
OPEs:
\begin{align}
\Psi_e(z) V_{\gamma;a}(w) \sim (z-w)^{l_{\gamma;a+1}} V_{\gamma; a +1} + \cdots,
\end{align}
where $V_{\gamma;a} = \Psi_e^a V_\gamma$ is a bound state of $a$ electrons and a quasiparticle.
The $\cdots$ indicate terms of order $\mathcal{O}( (z-w)^{l_{\gamma;a+1} + 1} )$,
The integer $n_{\gamma;l}$ is defined as the number of $a$ such that $l_{\gamma;a} = l$.
In the limit of large $l$, $n_{\gamma;l}$ is periodic and it is the unit cell that
characterizes a quasiparticle. For single-component states, these occupation number
sequences have been studied from many points of view and have proven to be an
important way of understanding the topological order of FQH states.\cite{WW0808,WW0809,BW0932,BW0937,
BH0802, BK0601,SL0604}

For $N \equiv p - q = 1$, the orbifold FQH phase is an Abelian
phase. The $Z_2$ vortices, which are non-Abelian excitations for 
$N >1$, have unit quantum dimension when $N = 1$ (see eqn. \ref{Z2vortexDeg}).
The ground state degeneracy on genus $g$ surfaces is
$[4(2p-1)]^g$, which shows that in fact all quasiparticles have unit quantum dimension.
Moreover, the $U(1)_2/Z_2$ orbifold CFT is actually equivalent to the $U(1)_8$
CFT,\cite{DV8985} which does not contain any primary operators with non-Abelian
fusion rules. 

Since this is an Abelian phase, it must exist within the $K$-matrix classification
of Abelian FQH phases.\cite{WZ9290} What is the $K$-matrix of the $N = 1$ orbifold states?
The $K$-matrix and charge vector $\v q$ are:
\begin{align}
\label{Kmatrix}
K = \left( \begin{matrix}
1 + q & q - 1 \\
q - 1 & 5 + q \\
\end{matrix} \right),
\;\;\;\;\;
\v q = \left(\begin{matrix}
1 \\
1\\
\end{matrix} \right).
\end{align}
In Section \ref{vertexAlg} we will explain how to arrive at this result. Notice
that this phase is actually a two-component bilayer state,
so we expect that the edge theory would contain two
electron operators, while in eqn. (\ref{elOp}) we only listed one electron operator. 
We will further explain this situation in Section \ref{vertexAlg} as well.

These $N = 1$ Abelian states are interesting because two-component Abelian states are 
all described by $U(1) \times U(1)$ CS theories:
\begin{align}
\mathcal{L} = \frac{1}{4\pi} K_{IJ} a^I \partial a^J + \frac{e}{2\pi} \v{q}_I A \partial a_I
\end{align}
Therefore, for the $K$-matrix in (\ref{Kmatrix}), we have found that there is 
a different yet equivalent Chern-Simons theory that describes the same phase. This 
other CS theory is the $U(1) \times U(1) \rtimes Z_2$ CS theory with the Lagrangian in
(\ref{lagrangian}) and with $p - q = 1$. 

\begin{table}
\begin{tabular}{clccc}
\hline
\multicolumn{5}{c}{Quasiparticles for $N = 2$ orbifold FQH state ($\nu = 1$)} \\
\hline
  & $Z_2$ Orbifold Label & $\{n_l\}$ & $h^{sc} + h^{ga}$ & q. dim.\\
\hline
$\v 0$ & $I$ &					 2  0 		& $0 + 0$  & 1\\
$\v 1$ & $\phi_N^1$ &					 0  2 		& $1/2 + 0$ & 1\\
&&&\\
$\v 2$ & $j$ &					 0  2 		& $1 + 0$  & 1 \\
$\v 3$ & $\phi_N^2$ &					 2  0 		& $1/2 + 0 = 1/2$ & 1 \\
&&&\\
$\v 4$ & $\sigma_1e^{i 1/2 \sqrt{\nu^{-1}} \phi_c}$ &		 1  1 		& $1/16 + 1/8$ & $\sqrt{2}$ \\
&&&\\
$\v 5$ & $\tau_1 e^{i 1/2 \sqrt{\nu^{-1}} \phi_c}$ &		 1  1 		& $9/16 + 1/8$ & $\sqrt{2}$ \\
&&&\\
$\v 6$ & $\sigma_2$ &                                  2  0 		& $1/16 + 0$ & $\sqrt{2}$ \\
$\v 7$ & $\tau_2$ &					 0  2 		& $9/16 + 0$ & $\sqrt{2}$ \\
&&&\\
$\v 8$ & $\cos(\frac{\varphi}{\sqrt{2}}) e^{i 1/2 \sqrt{\nu^{-1}}\phi_c}$ &		 1  1 		& $1/8 + 1/8$ & 2 \\
\hline
\end{tabular} 
\caption{
\label{N2qp}
Quasiparticle operators of the $(N,q) = (2,0)$ orbifold states, with 
filling fraction $\nu = 1$. The different representations of the
magnetic translation algebra\cite{BW0932} are separated by
spaces. $Q$ is the electric charge and $h^{sc}$
and $h^{ga}$ are the scaling dimensions of the 
orbifold primary field and the $U(1)$ vertex operator
$e^{i \alpha \varphi_c}$, respectively. $\varphi_c$ is a
free scalar boson that describes the charge sector.
$\{n_l\}$ is the occupation number sequence associated with
the quasiparticle pattern of zeros.
}
\end{table}

\begin{table}
\begin{tabular}{cllcc}
\hline
\multicolumn{5}{c}{Quasiparticles for $N=3$ orbifold state ($\nu = 2/3$)} \\
\hline
  & $Z_2$ Orbifold Label & $\{n_l\}$ & $h^{sc} + h^{ga}$ & q. dim.\\
\hline
$\v 0$ & $\mathbb{I} \sim \phi_N^1 e^{i \sqrt{3/2} \varphi_c} $ & 1 1 1 0 0 1 & 0+0 & 1\\
$\v 1$ & $e^{i 2/3 \sqrt{3/2} \varphi_c}$ & 1 1 1 1 0 0 & $0+\frac{1}{3}$ & 1\\
$\v 2$ & $e^{i 4/3 \sqrt{3/2} \varphi_c}$ & 0 1 1 1 1 0 & $0+\frac{4}{3}$ & 1\\
$\v 3$ & $j_r \sim \partial \varphi_r$ & 0 0 1 1 1 1 & 1+0 & 1\\
$\v 4$ & $j_r e^{i 2/3 \sqrt{3/2} \varphi_c}$ & 1 0 0 1 1 1 & $1+\frac{1}{3}$ & 1\\
$\v 5$ & $j_r e^{i 4/3 \sqrt{3/2} \varphi_c}$ & 1 1 0 0 1 1 & $1+\frac{4}{3}$ & 1\\
& & & & \\
$\v 6$ & $\sigma_1 e^{i 1/6 \sqrt{3/2} \varphi_c}$ & 1 1 0 1 0 1 & $\frac{1}{16} + \frac{1}{48}$ & $\sqrt{3}$\\
$\v 7$ & $\sigma_1 e^{i 5/6 \sqrt{3/2} \varphi_c}$ & 1 1 1 0 1 0 & $\frac{1}{16} + \frac{25}{48}$ & $\sqrt{3}$ \\
$\v 8$ & $\sigma_1 e^{i 9/6 \sqrt{3/2} \varphi_c}$ & 0 1 1 1 0 1 & $\frac{1}{16} + \frac{27}{16}$ & $\sqrt{3}$\\
$\v 9$ & $\tau_1 e^{i 1/6 \sqrt{3/2} \varphi_c}$ & 1 0 1 1 1 0 & $\frac{9}{16} + \frac{1}{48}$ & $\sqrt{3}$\\
$\v{10}$ & $\tau_1 e^{i 5/6 \sqrt{3/2} \varphi_c}$ & 0 1 0 1 1 1 & $\frac{9}{16} + \frac{25}{48}$ & $\sqrt{3}$\\
$\v{11}$ & $\tau_1 e^{i 9/6 \sqrt{3/2} \varphi_c}$ & 1 0 1 0 1 1 & $\frac{9}{16} + \frac{27}{16}$ & $\sqrt{3}$\\
& & & & \\
$\v{12}$ & $\cos(\frac{\varphi_r}{\sqrt{6}}) e^{i 1/3 \sqrt{3/2} \varphi_c}$  & 1 0 1 1 0 1 & $\frac{1}{12} + \frac{1}{12}$ & $2$\\
$\v{13}$ & $\cos(\frac{\varphi_r}{\sqrt{6}}) e^{i \sqrt{3/2} \varphi_c}$ & 1 1 0 1 1 0 & $\frac{1}{12} + \frac{3}{4}$ & $2$\\
$\v{14}$ & $\cos(\frac{\varphi_r}{\sqrt{6}}) e^{i 5/3 \sqrt{3/2} \varphi_c}$ & 0 1 1 0 1 1 & $\frac{1}{12} + \frac{25}{12}$ & $2$\\
\hline
\end{tabular}
\caption{\label{N3qp}
Quasiparticles in the $(N,q)= (3,0)$ orbifold FQH state, at 
$\nu = 2/3$. This state is also called the $Z_4$ parafermion
FQH state. The different representations of the
magnetic translation algebra\cite{BW0932} are separated by
spaces. $Q$ is the electric charge and $h^{sc}$
and $h^{ga}$ are the scaling dimensions of the 
orbifold primary field and the $U(1)$ vertex operator
$e^{i \alpha \varphi_c}$, respectively. $\varphi_c$ is a
free scalar boson that describes the charge sector.
$\{n_l\}$ is the occupation number sequence associated with
the quasiparticle pattern of zeros. }
\end{table}

\begin{table}
\begin{tabular}{clccc}
\hline
\multicolumn{5}{c}{Quasiparticles for $N = 4$ orbifold FQH state ($\nu = 1/2$)} \\
\hline
& CFT Label & $\{n_{\gamma;l}\}$ & $h^{sc} + h^{ga}$ & q. dim. \\ 
\hline
$\v 0$ & $\mathbb{I}$ &					 2  0  0  0 		& $0 + 0$ & 1\\
$\v 1$ & $e^{i 1/2 \sqrt{\nu^{-1}}\phi_c}$ &		 0  2  0  0 		& $0 + 1/4$ & 1\\
$\v 2$ & $\phi_N^1$ &					 0  0  2  0 		& $1 + 0 $ & 1 \\
$\v 3$ & $\phi_N^1e^{i 1/2 \sqrt{\nu^{-1}}\phi_c}$ &		 0  0  0  2 		& $1 + 1/4$ & 1\\
&&& \\
$\v 4$ & $j$ &					 0  1  0  1 		& $1 + 0$  & 1\\
$\v 5$ & $je^{i 1/2 \sqrt{\nu^{-1}}\phi_c}$ &		 1  0  1  0 		& $1 + 1/4$ & 1\\
$\v 6$ & $\phi_N^2$ &					 0  1  0  1 		& $1 + 0$ & 1\\
$\v 7$ & $\phi_N^2e^{i 1/2 \sqrt{\nu^{-1}}\phi_c}$ &		 1  0  1  0 		& $1 + 1/4$ & 1\\
&&&\\
$\v 8$ & $\sigma_1$ &					 0  1  0  1 		& $1/16 + 0$ & 2\\
$\v 9$ & $\sigma_1e^{i 1/2 \sqrt{\nu^{-1}}\phi_c}$ &		 1  0  1  0 		& $1/16 + 1/4$ & 2\\
\\
$\v{10}$ & $\tau_1$ &					 0  1  0  1 		& $9/16 + 0$ & 2\\
$\v {11}$ & $\tau_1e^{i 1/2 \sqrt{\nu^{-1}}\phi_c}$ &		 1  0  1  0 		& $9/16 + 1/4$ & 2\\
&&&\\
$\v{12}$ & $\sigma_2e^{i 1/4 \sqrt{\nu^{-1}}\phi_c}$ &		 1  1  0  0 		& $1/16 + 1/16$ & 2\\
$\v{13}$ & $\sigma_2e^{i 3/4 \sqrt{\nu^{-1}}\phi_c}$ &		 0  1  1  0 		& $1/16 + 9/16$ & 2\\
$\v{14}$ & $\tau_2e^{i 1/4 \sqrt{\nu^{-1}}\phi_c}$ &		 0  0  1  1 		& $9/16 + 1/16$ & 2\\
$\v{15}$ & $\tau_2e^{i 3/4 \sqrt{\nu^{-1}}\phi_c}$ &		 1  0  0  1 		& $9/16 + 9/16$ & 2\\
&&&\\
$\v{16}$ & $\cos(\frac{\varphi}{\sqrt{8}}) e^{i 1/4 \sqrt{\nu^{-1}}\phi_c}$ &		 1  1  0  0 		& $1/16 + 1/16$ & 2\\
$\v{17}$ & $\cos(\frac{\varphi}{\sqrt{8}})e^{i 3/4 \sqrt{\nu^{-1}}\phi_c}$ &		 0  1  1  0 		& $1/16 + 9/16$ & 2\\
$\v{18}$ & $\cos(\frac{3\varphi}{\sqrt{8}})e^{i 1/4 \sqrt{\nu^{-1}}\phi_c}$ &		 0  0  1  1 		& $9/16 + 1/16$ & 2\\
$\v{19}$ & $\cos(\frac{3\varphi}{\sqrt{8}})e^{i 3/4 \sqrt{\nu^{-1}}\phi_c}$ &		 1  0  0  1 		& $9/16 + 9/16$ & 2\\
&&&\\
$\v{20}$ & $\cos(\frac{2\varphi}{\sqrt{8}})$ &			         0  1  0  1 		& $1/4 + 0$ & 2\\
$\v{21}$ & $\cos(\frac{2\varphi}{\sqrt{8}})e^{i 1/2 \sqrt{\nu^{-1}}\phi_c}$ &		 1  0  1  0 		& $1/4 + 1/4$ & 2\\
\hline
\end{tabular}
\caption{
\label{N4qp}
Quasiparticles for the $(N,q) = (4,0)$ orbifold FQH states, at $\nu = 1/2$.
The different representations of the
magnetic translation algebra\cite{BW0932} are separated by
spaces. $Q$ is the electric charge and $h^{sc}$
and $h^{ga}$ are the scaling dimensions of the 
orbifold primary field and the $U(1)$ vertex operator
$e^{i \alpha \varphi_c}$, respectively. $\varphi_c$ is a
free scalar boson that describes the charge sector.
$\{n_l\}$ is the occupation number sequence associated with
the quasiparticle pattern of zeros.}
\end{table}

\section{Phase transition from orbifold FQH states to $(ppq)$ bilayer states}
\label{phaseTransition}

The phases described by the $U(1) \times U(1) \rtimes Z_2$ and 
$U(1) \times U(1)$ CS theories differ by an extra $Z_2$ gauge 
symmetry, which suggests that the transition between these
two phases is described by a $Z_2$ ``gauge symmetry-breaking'' 
transition. In this section we further elucidate this idea. 

First, consider the slave Ising construction presented in Section \ref{slaveSect}. 
There, we found that the difference between the $Z_2$ confined
and deconfined phases is associated with the condensation of
a $Z_2$ charged scalar field, $s^z$. When $\langle s^z_i \rangle = 0$,
the system is in the $Z_2$ deconfined phase and the low energy
theory is the $U(1) \times U(1) \rtimes Z_2$ CS theory. When
$\langle s^z_i \rangle \neq 0$, the low energy theory is the
$U(1) \times U(1)$ CS theory. This analysis suggests that these
two phases are separated by a continuous phase transition and that
the critical theory is simply a theory of the Ising field $s^z_i$
coupled to a $Z_2$ gauge field. This transition has been 
well-studied\cite{FS7982} and is known to be in the 3D Ising universality 
class. 

Now let us arrive at the above conclusion through totally different
arguments as well. From the CFT prescription for computing the
quasiparticle operators, we observe that the orbifold FQH states 
always contain an electrically neutral, topologically non-trivial 
Abelian quasiparticle with integer scaling dimension. In the
edge CFT, this quasiparticle is denoted $j \sim \partial \varphi_-$.
$j$ has trivial braiding properties with respect to itself because 
of its integer scaling dimension and is therefore a boson. 
It is another way of viewing the deconfined Ising spin $s^z$, so we 
expect it to carry $Z_2$ gauge charge. 
What happens when $j$ condenses? The condensation of
$j$ drives a topological phase transition to a state with 
different topological order. Based on general principles,\cite{BS0916}  
we can deduce that the topological order of the resulting 
phase is precisely that of the $(ppq)$ states.
This works as follows. 

Upon condensation, $j$ becomes identified with the identity 
sector of the topological phase. Any topologically inequivalent 
quasiparticles that differed from each other by fusion with $j$ 
will become topologically equivalent to each other after condensation. 
Furthermore, quasiparticles that were not local with respect to 
$j$ will not be present in the low energy spectrum of the theory 
after condensation. They become ``confined.'' Finally, if before 
condensation a quasiparticle $\gamma$ fused with its conjugate to 
both the identity and to $j$, then after condensation $\gamma$ 
splits into multiple topologically inequivalent quasiparticles. Otherwise, 
since $j$ is identified with the identity after condensation, there would be 
two ways for $\gamma$ to fuse with its conjugate to the identity, which is 
assumed to not be possible in a topological phase. 

Applying these principles, we can see that the condensation of 
$j$ yields the $(ppq)$ states. As a simple example, consider
the cases where $q = 0$. Some of the topological properties of the
$q = 0$ orbifold FQH states were described in Section \ref{genQPprop}.
When $j$ condenses, we find that $A_l$ becomes topologically identified
with $A_{l+N}$. The quasiparticles labelled by $B_l$ become confined, because
the OPE of the operator $j$ with the operators $B_l$ in the edge theory
always have a branch cut and so the $B_l$ are nonlocal with 
respect to $j$. Finally, the quasiparticles $C_{mn}$
each split into two distinct quasiparticles. This leaves a total of 
$N^2$ Abelian quasiparticles whose topological properties all agree
exactly with those of the $(NN0)$ states. 

From the results of the $U(1) \times U(1) \rtimes Z_2$ CS theory, 
we find that $j$ carries $Z_2$ gauge charge, proving that it is 
indeed associated with $s^z$ in the slave Ising description. We arrive at this 
result by first studying the number of $Z_2$ non-invariant states 
in the presence of $n$ pairs $Z_2$ vortices at fixed locations on 
a sphere (see eqn. (\ref{Z2noninv}) ). We observe that this number 
coincides exactly with the number of ways for $n$ pairs of the 
fundamental non-Abelian quasiparticles and their conjugates to 
fuse to $j$. That is, we can use the CFT prescription to calculate 
the fusion rules
\begin{align}
(B_l \times \bar{B}_l)^n = a_n \mathbb{I} + b_n j + \cdots,
\end{align}
and we observe that $b_n$ agrees exactly with the $\beta_n$ in
eqn. (\ref{Z2noninv}). This shows that $j$ carries $Z_2$ 
gauge charge. This makes sense from the perspective of the low energy
theory, because the condensation of $j$ yields 
a Higgs phase of the $Z_2$ sector and leaves us with the 
$U(1) \times U(1)$ CS theory, which describes the $(ppq)$ states.
Moreover, in the edge theory, $j \sim \partial \varphi_-$ is odd under the $Z_2$ 
transformation $\varphi_- \rightarrow -\varphi_-$, which is consistent with
the fact that $j$ carries the $Z_2$ gauge charge in the bulk; the
$Z_2$ in the orbifold sector of the edge theory is the ``same'' $Z_2$
as the $Z_2$ gauge transformation that interchanges the two $U(1)$
gauge fields in the $U(1) \times U(1) \rtimes Z_2$ CS theory. 

As the energy gap to creating excitations of $j$ is reduced 
to zero, the low energy theory near the transition must be that 
of a $Z_2$ gauged Ginzburg-Landau theory and the transition is 
therefore in the 3D Ising universality class.\cite{BW102,FS7982}

This close connection between the topological properties of the orbifold FQH states 
and those of the bilayer $(ppq)$ states provides additional strong evidence for why the 
CFT prescription presented in Section \ref{edgeSec} is correct and describes the same 
topological theory as the $U(1) \times U(1) \rtimes Z_2$ CS theory. 
From the $U(1) \times U(1) \rtimes Z_2$ CS theory, we know that there must be 
a $Z_2$ Higgs transition to the $(ppq)$ states, and so the topological 
quantum numbers of the orbifold phase must agree with the
condensate-induced transition to the $(ppq)$ states. The CFT prescription presented in 
Section \ref{edgeSec} provides us with such a consistent set of topological quantum
numbers. 

We note that while the edge between the orbifold FQH states and a topologically
trivial phase will have protected edge modes, we do not expect protected edge modes
at the edge between the orbifold states and the corresponding $(ppq)$ states, because
they differ by a $Z_2$ transition.\footnote{We thank Michael Levin for conversations
regarding this point.} As a simple check, note that the edge CFT for both states has
central charge $c = 2$, so the edge between these two states would have zero thermal
Hall conductance.

\subsection*{Anyon condensation and transition from $(ppq)$ states to orbifold FQH states}
\label{anyonCond}

The above discussion showed that we may understand the transition from
the orbifold FQH states to the $(ppq)$ states through the condensation
of an electrically neutral boson, ultimately leading us to conclude that the
transition is continuous and is in the 3D Ising universality class. 

An interesting, though currently unresolved, question relates to
how we should understand this phase transition from the other direction:
starting from the $(ppq)$ states and ending with the orbifold FQH states. 
Perhaps this transition can be understood as the condensation of
the Abelian anyons of the $(ppq)$ state into some collective state
and driving a phase transition to a more complicated topological phase. 

Starting from the $(ppq)$ state, which can exist in the absence of interlayer
tunneling, we expect that the orbifold FQH state can be stabilized for some
range of interlayer tunneling. There are two reasons for this expectation. 
First, in the slave Ising construction 
we see that the fluctuations of the $Z_2$ charged boson are accompanied by interlayer density
fluctuations. We can see this relation by noticing that the relative density difference
between layers is (see Section \ref{slaveSect}):
\begin{align}
n_r &= \Psi_{e\uparrow}^\dagger \Psi_{e\uparrow} - \Psi_{e\downarrow}^{\dagger} \Psi_{e\downarrow} 
\nonumber \\
&= \Psi_{e+}^{\dagger} \Psi_{e-} + \Psi_{e-}^{\dagger}\Psi_{e+}  = s^z (c_+^{\dagger} c_- + c_-^{\dagger} c_+).
\end{align}
In the slave particle gauge theory, the states associated with the $c$ fermions do not 
change as we tune through the transition, while the fluctuations of $s^z$ become critical
at the transition, thus leading to interlayer density fluctuations as well.  Since the interlayer density fluctuations
can physically be tuned by the interlayer tunneling, we expect that interlayer tunneling will
be one of the material parameters that will help tune through this transition.

A second, closely related, consideration that suggests interlayer tunneling can tune through this transition 
is the following. In the absence of electron tunneling, we have a bilayer state with two electron operators, 
$\Psi_{e\uparrow}$ and $\Psi_{e\downarrow}$, for the two layers. As the interlayer tunneling is increased, there
will be a single-particle gap between the symmetric and anti-symmetric orbitals that also increases. In the limit
of large interlayer tunneling, we expect all of the electrons to occupy the symmetric orbitals; thus, the 
electron operator that we need to be concerned with in the limit of large interlayer tunneling is
$\Psi_{e+} \propto \Psi_{e\uparrow} + \Psi_{e\downarrow}$. Now, the $(NN0)$ states can be obtained from 
a parton construction by setting
\begin{align}
\Psi_{e\uparrow} &= \psi_1 \cdots \psi_N,
\nonumber \\
\Psi_{e\downarrow} &= \psi_{N+1} \cdots \psi_{2N},
\end{align}
where $\psi_i$ are charged $e/N$ fermions that form $\nu = 1$ IQH states. The gauge group here is
$SU(N) \times SU(N)$, and integrating out the partons gives $SU(N)_1 \times SU(N)_1$ CS theory. 
In the phase where the interlayer tunneling is high, so that we effectively have a single electron operator
$\Psi_{e} = \Psi_{e+} =  \frac{1}{\sqrt{2}} (\psi_1 \cdots \psi_N + \psi_{N+1} \cdots \psi_{2N})$, and
we ignore the other electron operator completely, the gauge group for 
$N > 2$ is $SU(N) \times SU(N) \rtimes Z_2$, which after integrating out the
partons is equivalent to the $U(1) \times U(1) \rtimes Z_2$ CS theory for the orbifold states. 

The above two considerations suggest that interlayer tunneling will allow us to tune through
the phase transition. Since the important dimensionless parameters are $t/V_{inter}$ and 
$V_{inter}/V_{intra}$, where $V_{inter/intra}$ are the inter/intra-layer
Coulomb repulsions and $t$ is the interlayer tunneling, we expect that 
tuning the Coulomb repulsions should also be able to stabilize the orbifold state. 
Now, as discussed in \Ref{W0050, BW102}, observe that bilayer FQH states 
have a particular kind of neutral excitation called a fractional exciton (f-exciton), 
which is the bound state of a quasiparticle in one layer and a quasihole in 
the other layer. The f-exciton carries fractional statistics;
when the interlayer Coulomb repulsion is increased, the gap to the f-exciton can be 
decreased to zero. This means that interlayer repulsion can drive a phase transition
between two fractional quantum Hall states at the same filling fraction. This leads
us to suspect that this anyon condensation can be related to the formation of the orbifold 
FQH states, because the orbifold states can also be obtained by tuning the Coulomb repulsions
in the presence of interlayer tunneling. Of course, anyons can condense in different ways,
and likewise we expect that there will be other microscopic interactions that will determine
precisely which phases will be obtained when the interlayer tunneling/repulsion are tuned.

\section{Ideal wave functions and the vertex algebra approach to the orbifold FQH states}
\label{vertexAlg}

In the sections above, we have introduced and developed a theory for a novel
series of non-Abelian FQH states: the orbifold FQH states. These are parameterized
by two integers, $(N,q)$. They occur at filling fraction $\nu = 2/(N + 2q)$
and are separated from the $(ppq)$ states (where $N = p-q$) by a continuous $Z_2$ 
phase transition. 

For $N = 3$, these states are equivalent to the $Z_4$ parafermion states, which
have an ideal wave function description.\cite{RR9984} In other words,
if we take the electron operator in (\ref{elOp}), and evaluate the following correlator:
\begin{align}
\Psi(\{z_i\}) \sim \langle V_e (z_1) \cdots V_e(z_N) \rangle,
\end{align}
we will obtain a wave function that describes an incompressible FQH state. However,
carrying out this construction for $N \neq 3$ will not yield a wave function
that describes an incompressible FQH state. In fact, for $N > 3$, the pattern
of zeros of the electron operator $V_e$ corresponds to certain problematic, or sick,
pattern of zeros solutions: pattern of zeros solutions whose relevance to
describing gapped topological phases had been uncertain because their associated
ideal wave functions always appear to be gapless.\cite{LW1024} 

In the following, we will study the orbifold FQH states from the pattern of zeros
and ideal wave function point of view. The main conclusion to draw is that
the sick pattern of zeros solutions are still relevant to quantitatively 
characterizing topological order in FQH states, even when naively it appears as 
though the corresponding ideal wave function is gapless! In the analysis below, 
we will see how the orbifold FQH states provide profound lessons for the conceptual
foundation of the pattern of zeros/vertex algebra approach to constructing 
ideal wave functions. 

\subsection{Review of the vertex algebra/conformal field theory approach}

A wide class of FQH states can be described by ideal wave functions that are 
exact zero-energy ground states of Hamiltonians with interactions that are
either delta functions or derivatives of delta functions. Such ideal Hamiltonians 
select for certain properties of the ground state wave functions, 
such as the order of the zeros in the wave function as various numbers of 
particles approach each other. 

The ideal wave functions that we currently understand can all 
be written in terms of a correlation function of vertex operators:
\begin{align}
\label{idealwfn}
\Psi(\{z_i\}) \sim \langle V_e(z_1) \cdots V_e(z_N) \rangle, 
\end{align}
where $V_e$ is a certain operator in a 2D chiral CFT, called the electron operator. 
The wave function $\Psi(\{z_i\})$ can be specified by simply specifying the 
operator algebra generated by the electron operator:
\begin{align}
V_e(z) V_e(w) &\sim C_{ee\mathcal{O}_1}(z-w)^{h_{\mathcal{O}_1} - 2h_e} \mathcal{O}_1 + \cdots
\nonumber \\
V_e(z) \mathcal{O}_1(w) &\sim C_{e\mathcal{O}_1\mathcal{O}_2} (z-w)^{h_{\mathcal{O}_2}- h_e - h_{\mathcal{O}_1}} \mathcal{O}_2 + \cdots
\nonumber \\
\vdots
\end{align}
This operator algebra is called a vertex algebra. Using this vertex algebra, 
the correlation function (\ref{idealwfn}) can be evaluated by successively 
replacing products of two neighboring operators by a sum of single operators. 
In order for the result to be independent of the order in which these successive 
fusions are evaluated, there need to be various consistency conditions
on the vertex algebra. In some cases, specifying the scaling dimension $h_e$  and the
filling fraction $\nu$ is enough to completely specify the vertex algebra, because the
structure constants $C_{abc}$ can be obtained through the various consistency conditions.\cite{ZF8515}
In these cases, an ideal Hamiltonian that selects for the pattern of zeros 
is believed to have a unique zero energy wave function of highest density. Otherwise, 
one needs to find a way to use the Hamiltonian to select also for a certain choice of 
structure constants $C_{abc}$. 

The quasiparticle wave functions can also be written as correlators:
\begin{align}
\label{qpwfn}
\Psi_\gamma(\eta; \{z_i\}) \sim \langle V_{\gamma} (\eta) V_e(z_1) \cdots V_e(z_n)\rangle, 
\end{align}
where $V_\gamma$ is a ``quasiparticle'' operator and $\eta$ is the position of the
quasiparticle. To evaluate these wave functions, we need to specify the
operator algebra involving the quasiparticle operators. In order for the quasiparticle
wave function (\ref{qpwfn}) to be single-valued in the electron coordinates, 
the allowed quasiparticle operators must be local with respect to the electron operators --
their operator product expansion with the electron operator must not contain any branch cuts. 
Two quasiparticle operators are topologically equivalent if they are related 
by electron operators. By solving the consistency conditions on the vertex algebra, 
we can obtain the constraints on the allowed quasiparticles. In the vertex algebra approach
to FQH states, we take all solutions of the consistency conditions to be 
valid quasiparticle operators, so there can be a finite number of quasiparticles only 
if the number of solutions to the consistency conditions is finite. This is equivalent
to the expectation that ideal Hamiltonians cannot selectively pick some of the quasiparticle
vertex algebra solutions as allowed zero-energy states and not others. This expectation is
fulfilled in all known FQH states that can be described by ideal Hamiltonians and ideal
wave functions. 

When the ideal Hamiltonian can uniquely select for the zero
energy wave function of highest density, when there are
a finite number of solutions to the quasiparticle
consistency conditions, and when the vertex algebra is unitary\cite{R0908}
we believe that these model wave
functions belong to an incompressible FQH phase. Its
topological properties are dictated by the properties of the
quasiparticle operators in the CFT. Such is the case for the
Read-Rezayi states and some of their generalizations.\cite{MR9162,RR9984,SR0718,AS9996,AL0205,WW0808}
Remarkably, it is also the case that the edge CFT is the
same as the CFT whose correlation function yields the ideal
wave function. 

For some other choices of vertex algebra, 
there are an infinite number of solutions to the
associativity conditions for a quasiparticle.\cite{LW1024}
Such a situation means that the corresponding ideal wave
function likely does not describe a gapped phase.

The orbifold FQH states are interesting because if we try to
use their edge CFT to construct single-component ideal wave
functions, we find that for $N > 3$, the corresponding ideal
wave function is gapless.  The vertex algebra of the
electron operator allows for an infinite number of
quasiparticle solutions, indicating the gapless nature of
the ideal wave function.\cite{LW1024} The case $N = 3$ is
special: the pattern of zeros of the electron operator
uniquely fixes the ground state wave function and there are
a finite number of quasiparticle solutions for the vertex
algebra -- this corresponds to the $Z_4$ parafermion FQH
states and it possesses a well-behaved single-layer ideal
wave function. For $N = 1,2$, we find that the single-layer
wave function is gapped but does not have the topological
properties of the orbifold states; in order to have a
description of these states in terms of ideal wave functions
we are forced to view the orbifold FQH states as
double-layer states.

In order to shed light on the pattern of zeros/vertex
algebra approach to constructing FQH states, we study the
orbifold FQH states from this point of view. The analysis
below suggests that while some of the apparently sick
pattern of zeros/vertex algebra solutions may not describe
gapped FQH phases, they lie near a critical point and can be
driven to a nearby incompressible phase -- the orbifold FQH
state -- by applying certain perturbations to the ideal
Hamiltonian. In the vertex algebra framework, this
corresponds to enlarging the vertex algebra by incorporating
additional local operators. 

\subsection{Orbifold FQH states viewed through vertex algebra}

In Section \ref{edgeSec} we explained that the electron operator in the orbifold
FQH edge theory is given by the operator
\begin{align}
V_e(z) = \phi^1_N(z) e^{i \sqrt{\nu^{-1}} \phi_c}, 
\end{align}
where $\phi^1_N$ is an operator from the $U(1)_{2N}/Z_2$ CFT
and has scaling dimension $h_{\phi^1_N} = N/4$. When $N$ is
even, we have the following fusion rule
\begin{align}
\phi^1_N \times \phi^1_N = \mathbb{I}.
\end{align}
When $N$ is odd, we have
\begin{align}
\phi^1_N \times \phi^1_N = j, \;\;\;\; j \times j = \mathbb{I}.
\end{align}
These fusion rules denote fusion between representations of
the orbifold chiral algebra $\mathcal{A_N}/Z_2$. The identity
representation still contains an infinite set of Virasoro representations,
labelled by the Virasoro primary fields $\cos(l \sqrt{2N} \varphi)$, for integer $l$.

Our task will be to study the pattern of zeros of these electron operators,
$V_e$, compare with results from the pattern of zeros approach\cite{WW0808,WW0809,BW0932,BW0937} 
and with the vertex algebra approach,\cite{LW1024}
and try to make sense of any discrepancies. Since the discussion depends
on the choice of $N$, we will study various choices of $N$ individually. 
We note that the pattern of zeros that we calculate from the electron operator,
using the prescription of Section \ref{edgeSec},
assumes that the highest weight field appears in the OPEs if they
are allowed by the fusion rules. In other words, the structure constants
involving the highest weight fields are assumed to be nonzero. This is 
consistent with cases in which the $Z_2$ orbifold vertex algebra is known 
(e.g. for $N = 3$ because of the relation to $Z_4$ parafermion CFT), 
and can perhaps be viewed as a consequence of the naturality theorem for rational 
CFTs.\cite{MS8916}


\subsubsection{$N = 1$}
\label{N1Orb}

Here the electron operator is given by:
\begin{align}
V_e(z) = \phi_N^1 e^{i \sqrt{1/2 + q} \phi_c(z)},
\end{align}
and $\phi_N^1$ has scaling dimension $1/4$. The pattern of zeros
associated with $V_e(z)$ is the pattern of zeros of the
Laughlin $\nu  = 1/(q+1)$ wave function, which describes a state
with a different topological order than the $N = 1$ orbifold FQH states. 
An ideal Hamiltonian that selects for such a pattern of zeros will 
actually yield the Laughlin $\nu = 1/(q+1)$ state and not the
$N = 1$ orbifold FQH state. 

In order to obtain an ideal wave function for the orbifold FQH state, we need to reinterpret the
system as a bilayer system. This means that we need to specify a second
electron operator. The second electron operator will 
resolve the difference between $e^{i \sqrt{q + 1} \phi_c(z)}$ -- whose correlation function 
yields the Laughlin states -- and $V_e(z) = \phi^1_N e^{i \sqrt{1/2 + q} \phi_c(z)}$,
because these two operators will have a different pattern of zeros
as viewed by the second electron operator. Another way to think about this
is in terms of the ideal Hamiltonian. When the Hilbert space is enlarged
to that of a double-layer system, the original ideal Hamiltonian -- which
only selects for the way wave functions go to zero when one flavor of particles
comes together -- will be gapless. It can be modified by adding additional terms
that also select for the pattern of zeros involving the other flavor of particles.
This modified Hamiltonian may then be gapped. 

Returning to the vertex algebra language, notice that it suffices to 
add an electrically neutral bosonic operator $V_{o}$ to the chiral algebra; 
then the composite operator $V_e V_o$ will be considered
as the second electron operator. In order to do this, it is helpful to observe
that the $U(1)_2/Z_2$ CFT is actually dual to the $U(1)_8$ CFT, whose chiral algebra 
is generated by the operators $e^{\pm i\sqrt{8} \phi_n(z)}$, where $\phi_n$
is a free chiral boson. The operators $\phi^1_N$ and $\phi_N^2$ are then equivalent to the 
operators $e^{\pm i \phi_n/\sqrt{2}}$, both of which have scaling dimension $1/4$. 
This suggests that we should seek an operator of the form
\begin{align}
V_o = e^{i l \sqrt{2} \phi_n},
\end{align}
because for any integer $l$ it is local with respect to 
$e^{\pm i \phi_n/\sqrt{2}} e^{i \sqrt{1/2 + q} \phi_c}$ and it is bosonic. 

For each $l$, we can design an ideal bilayer Hamiltonian so that the bilayer
wave function
\begin{align}
\Psi(\{z_i\}, \{w_i\}) \sim \langle \prod_{i} V_{e1}(z_i)V_{e2}(w_i) \rangle
\end{align}
is an exact zero energy ground state and the unique one of highest density. 
Here, $V_{e1} = e^{i \frac{1}{\sqrt{2}} \phi_n(z)} e^{i \sqrt{1/2 + q} \phi_c(z)}$
and $V_{e2} = V_{e1} V_o = e^{i \frac{2l+1}{\sqrt{2}} \phi_n(z)} e^{i \sqrt{1/2 + q} \phi_c(z)}$.
These states correspond to bilayer Abelian states with $K$-matrix and charge
vector 
\begin{align}
K = \left( \begin{matrix}
q+1 & q+l+1 \\
q+l+1 & q+1+2l+l^2 \\
\end{matrix} \right),
\;\;\;\;\;
\v q = \left( \begin{matrix}
1 \\
1 \\
\end{matrix} \right)
\end{align}

The case $l = 1$ corresponds to the $(ppq)$ states, while the case
$l = 2$ corresponds to the orbifold FQH states with $N = 1$. 
We comment on other choices of $l$ in Appendix \ref{ZNtrans}.

The $l = 1$ $(ppq)$ state and the $l = 2$ orbifold state
with $N = 1$ are connected by a continuous phase transition.
In the $l = 2$ orbifold state, the operator $e^{i {\sqrt{2}}
\phi_n}$ is a topologically non-trivial neutral boson that
squares to $V_{o}$, which lives in the identity sector.
When this neutral boson $e^{i {\sqrt{2}} \phi_n}$ condenses,
it is added to the identity sector and we obtain the $(ppq)$
states. 

Therefore, we see that the original gapless ideal
Hamiltonian can be perturbed to many different
incompressible phases. The critical point contains many
different bosons that can be condensed; condensing a
particular one will yield a particular gapped FQH state.
From the vertex algebra point of view, there are many
different bosonic operators that can be added to the vertex
algebra. One particular choice ($l=2$) will yield the $N =
1$ orbifold states, while another choice ($l=1$) will yield
the $(ppq)$ states.

\subsubsection{$N = 2$} 

The case $N = 2$ is similar to the case $N = 1$ in that these
orbifold states also need to be interpreted as bilayer states
in order for the ideal Hamiltonian to yield the orbifold FQH phase.
If we take the electron operator
\begin{align}
V_e(z) =  \phi_N^1(z) e^{i \sqrt{q+1} \phi(z)}
\end{align}
for $N = 2$, then we see that it has the same pattern of zeros
as the Pfaffian ground state wave function at $\nu = 1/(q+1)$ (see e.g. Table \ref{N2qp}).
In order to construct an ideal wave function for the orbifold 
FQH phase, we need to reinterpret the system as a two-component phase, 
which again means adding a second electron operator to the chiral algebra.
We leave a detailed analysis of this for future work.

\subsubsection{$N = 3$}

For $N = 3$, we find that the electron operator
\begin{align}
V_e(z) =  \phi_N^1(z) e^{i \sqrt{3/2 + q} \phi(z)}
\end{align}
has the same pattern of zeros as the $Z_4$ parafermion wave functions,
which are known to be exact ground states of single-layer ideal Hamiltonians. 
The topological order of the $Z_4$ parafermion states is that of the 
orbifold states with $N = 3$. Thus for $N = 3$, the ideal wave functions 
and ideal Hamiltonians do properly describe 
the orbifold phases. 

\subsubsection{$N = 4$}

The $N = 4$ case is the first highly non-trivial example that we encounter.
The pattern of zeros of the electron operator
\begin{align}
\label{N4elOpOrb}
V_e(z) =  \phi_N^1(z) e^{i \sqrt{2 + q} \phi(z)}
\end{align}
corresponds to the pattern of zeros associated with multiplying
a $\nu = 1$ Pfaffian wave function by a $\nu = 1/(q+1)$ Pfaffian
wave function. For $q = 0$, the pattern of zeros is simply that of
the square of the $\nu = 1$ Pfaffian wave function, which
is called the Haffnian wave function:
\begin{align}
\Phi_{Haffnian} &= \left( Pf \left( \frac{1}{z_i - z_j} \right) \right)^2 \prod_{i < j} (z_i - z_j)^2
\nonumber \\
&= \mathcal{S} \left( \frac{1}{(z_i-z_j)^2} \right) \prod_{i < j} (z_i - z_j)^2.
\end{align}
Here, $\mathcal{S}$ denotes symmetrization: 
$\mathcal{S} (M_{ij}) = \sum_{P} M_{P(1)P(2)} \cdots M_{P(N_e-1)P(N_e)}$, where $\sum_P$ is
the sum over all permutations of $N_e$ elements.
This pattern of zeros was studied in detail through the vertex algebra framework
in \Ref{LW1024}; the vertex algebra there was named $Z_2 | Z_2$ vertex algebra. 
It was found that 
the structure constants for one class of quasiparticles come with a 
free continuous parameter, indicating that the ideal wave function is likely
gapless (see also \Ref{R0908} for a similar discussion). 
This conclusion of gaplessness is corroborated from a totally 
different analysis:\cite{G01} the Haffnian wave function corresponds to
a critical point of $d$-wave paired composite fermions.

However, the $N = 4$ orbifold FQH states indeed exist as gapped FQH states,
and in particular from Table \ref{N4qp} we see that many of the quasiparticle
pattern of zeros are repeated -- there is not a one-to-one correspondence
between the pattern of zeros and the topologically distinct quasiparticles. 
In the following, we describe how to understand these results through
the vertex algebra framework.

From \Ref{LW1024} we learn that one set of sequences $\{n_{\gamma;l}\}$
are associated with operators whose structure constants can take on any
continuous parameter, $\alpha$. For certain discrete values of $\alpha$,
the associated quasiparticle is a boson, \it ie \rm it is local with respect
to itself, and may also be local with respect to the electron operator.
In such a case, this operator can and should be added to the chiral vertex
algebra and treated as a second electron operator. From the point of view
of the ideal Hamiltonian, this is like adding a perturbation so that the
system is driven away from the critical point and into a nearby incompressible phase.
The perturbation should be viewed as driving the condensation of a bosonic 
operator, which adds a second component to the FQH state. 

Since this nearby incompressible phase should now be viewed as a two-component state,
it thus should have an ideal wave function description in terms
of a double-layer state, described by the enlarged chiral algebra. Note 
that there may be several different, mutually exclusive choices 
for which operator to add to the chiral algebra; equivalently, 
there may be several different directions in which to perturb
the ideal Hamiltonian, each of which leads to a different incompressible phase.
When the chiral algebra is enlarged in such a way, then the continuous set of quasiparticles described by $\alpha$
will not all be allowed and instead only a finite, discrete set of them will be distinct and
local with respect to both the original electron operator and the second electron
operator. From this point of view we can now understand why the pattern of zeros
for the other quasiparticles may appear multiple times, even when there was a unique
solution for the structure constants: there is now an additional electron operator
that may resolve the difference between two quasiparticles that the first electron 
operator could not distinguish between. 

Let us study this more concretely for the case $q = 0$ using the vertex algebra 
framework. The vertex algebra generated by the electron operator (\ref{N4elOpOrb}) 
contains two pieces, the ``charge'' part, described by the $U(1)$ vertex operator
$e^{i \sqrt{2} \phi_c}$, and the ``neutral'' part, described by the operator 
$\phi_N^1$. For $N = 4$, $\phi_N^1$ is an operator with unit scaling dimension, and
its vertex algebra is equivalent to the vertex algebra of the current operator
$j \sim \partial \varphi$ in a free boson theory. Thus as far as the vertex
algebra of the electron operator is concerned, the electron operator may
be written as
\begin{align}
\label{elOpN4VA}
V_e = j e^{i \sqrt{2} \phi_c}.
\end{align}
In order for an operator $\mathcal{O}$ to be an allowed representation of 
the vertex algebra, it has to satisfy various consistency conditions, such as the
generalized Jacobi identity. In \Ref{LW1024}, the allowed representations of 
the algebra generated by $j$ were systematically studied (without embedding it into
a free boson theory). It was found that there is one operator, 
$\sigma$, with scaling dimension $1/16$, and another
continuous set of quasiparticles with continuously varying scaling dimensions. 
These operators are familiar when the algebra is embedded into a free boson
theory: $\sigma$ is the well-known twist operator, whose insertion at a point
in space induces a branch cut around which $\varphi \rightarrow -\varphi$. 
The continuously varying set of operators correspond to the operators
$e^{i\alpha \varphi}$. We see that $\varphi$ is uncompactified, because all
operators of the form $e^{i \alpha \varphi}$ are local with respect to 
$j$ and therefore correspond to distinct, allowed representations of the 
electron chiral algebra. 

Now notice that $e^{i \alpha \varphi}$ is bosonic when it has integer 
scaling dimension: $h_\alpha = \alpha^2/2 \in \mathbb{Z}$. Suppose that
we add the bosonic operator $\cos(\sqrt{8} \varphi)$ to the vertex algebra. 
Intuitively, we expect that this will cause only a discrete set of
the operators $e^{i\alpha \varphi}$ to now be local with respect to the enlarged
vertex algebra, and only a finite set of them will correspond to distinct
representations of the algebra. Put another way, inclusion of $\cos(\sqrt{8} \varphi)$
into the vertex algebra has the effect of essentially compactifying the boson $\varphi$, which then
quantizes the possible values of $\alpha$. To find all of the allowed, distinct 
quasiparticles, we want to find all solutions to the consistency 
conditions\cite{LW1024} for allowed quasiparticle operators for this enlarged
algebra. Here, we will not solve these conditions and instead save this
analysis for future work. Instead, we will verify this picture semi-rigorously
using the following arguments.

The operator $\cos(\sqrt{8} \varphi)$ is known to generate the 
chiral algebra of the $U(1)_8/Z_2$ orbifold CFT. The operator content
of the $U(1)_8/Z_2$ CFT is known, so we will assume that such an
operator content is in one-to-one correspondence with the distinct, 
allowed operators that satisfy the consistency conditions of the vertex
algebra generated by $\cos(\sqrt{8} \varphi)$. This is a reasonable
assumption, however  it has not been verified because the operator
content of the $U(1)/Z_2$ orbifold CFT was derived using considerations of modular
invariance, and not through directly studying the representation theory
of the chiral algebra by solving consistency conditions. For example, 
the $U(1)/Z_2$ CFT contains two twist fields $\sigma_1$ and $\sigma_2$ 
(and their counterparts $\tau_1$ and $\tau_2$, where $j \times \sigma_i = \tau_i$). 
The fact that there must be two twist fields can be understood through 
considerations of modular invariance of the torus partition function,
which includes both holomorphic and anti-holomorphic sectors of the theory.
That there should be two sets of twist fields is less obvious from the 
point of view of solving consistency conditions of the vertex algebra. 

In light of the above assumption, it is now clear what we should do. 
The allowed quasiparticles will be of the form
\begin{align}
\label{qpOpVAN4}
V_{qp} = \mathcal{O}e^{iQ \sqrt{2} \phi_c},
\end{align}
where $\mathcal{O}$ is a primary operator from the $U(1)_{2N}/Z_2$ orbifold theory.
We need to find all possible operators $V_{qp}$ that are local with respect
to the operators $V_e = j e^{i \sqrt{2} \phi_c}$, and $\cos(\sqrt{8} \varphi)$.
All operators of the form (\ref{qpOpVAN4}) are already local with respect to
$\cos(\sqrt{8} \varphi)$ because they form representations of the algebra
generated by $\cos(\sqrt{8} \varphi)$, so we only need to worry about their being local
with respect to $V_e = j e^{i \sqrt{2} \phi_c}$. Remarkably, carrying 
this out yields all of the topological properties of the
$N = 4$ orbifold states! In particular, we obtain the same results as we did using the
prescription used earlier in Sections \ref{edgeSec} and \ref{QPcontent} 
(see Table \ref{Z2orbQP} and compare with Table \ref{N4qp}).
Note that while the CFT labelling of the operators is different, 
the two prescriptions yield exactly the same topological
properties. This agreement is highly non-trivial and only works for 
$N = 4$, because only for $N = 4$ is the algebra of $j e^{i\sqrt{2} \phi_c}$ 
the same as the algebra of $\phi_N^1 e^{i \sqrt{2} \phi_c}$.
 
This adds evidence to the picture presented here, where the orbifold 
FQH states can be interpreted through the vertex algebra language as though
an additional bosonic operator has been added to the chiral algebra. 
In order to more rigorously confirm this picture, we would need to systematically solve 
the consistency conditions on the vertex algebra generated by $j e^{i\sqrt{2} \phi_c}$
and $\cos(\sqrt{8} \varphi)$ and show that the quasiparticle solutions and their
properties coincide with those presented here. 

Note that since we now have two electron operators, the full pattern of zeros
characterization should be described by the sequence $\{S_{\vec a}\}$, where
$\vec a$ is now a two-dimensional vector.\cite{BW0937} Therefore, 
the results of Table \ref{Z2orbQP} do not display this full pattern of zeros/vertex 
algebra data and instead only display the pattern of zeros as seen by the
electron operator $j e^{i\sqrt{2} \phi_c}$.

\begin{table}
\begin{tabular}{llccc}
\hline
\multicolumn{5}{c}{$U(1)_{2N}/Z_2$ Orbifold FQH states, with $V_e = j e^{i \sqrt{2} \varphi_c}$}\\
\hline
 & CFT Operator & $\{n_l\}$  & $h^{sc} + h^{ga}$ & Quantum Dim.\\
\hline
$\v 0$ & $\mathbb{I}$ & 2 0 0 0 & 0 + 0 & 1\\
$\v 1$ & $e^{i 1/2 \sqrt{2} \varphi_c}$ & 0 2 0 0  & 0 + 1/4 & 1 \\
$\v 2$ & $ e^{i \sqrt{2} \varphi_c} $  & 0 0 2 0 & 0 + 1 & 1\\
$\v 3$ & $e^{i 3/2 \sqrt{2} \varphi_c}$ & 0 0 0 2 & 0 + 9/4 & 1\\
\\
$\v 4$ & $\phi_N^1$ & 0 1 0 1 & $1+0$ & $1$ \\
$\v 5$ & $\phi_N^1 e^{i 1/2 \sqrt{2} \varphi_c}$ & 1 0 1 0  & $1 + 1/4$ & $1$\\
$\v 6$ & $\phi_N^2 $ & 0 1 0 1 & $1+0$ & $1$ \\
$\v 7$ & $\phi_N^2 e^{i 1/2 \sqrt{2} \varphi_c}$ & 1 0 1 0 & $1 + 1/4$ & $1$ \\
\\
$\v 9$ & $\phi_1$ & 0 1 0 1 & $1/16 + 0$ & $2$ \\
$\v 8$ & $\phi_1 e^{i 1/2 \sqrt{2} \varphi_c}$ & 1 0 1 0 & $1/16 + 1/4$ & $2$ \\
\\
$\v{10}$ & $\phi_3$ & 0 1 0 1 & $9/16 + 0$ & $2$ \\
$\v{11}$ & $\phi_3 e^{i 1/2 \sqrt{2} \varphi_c}$ & 1 0 1 0 & $9/16 + 1/4$ & $2$ \\
\\
$\v{12}$ & $\sigma_1 e^{i 1/4 \sqrt{2} \varphi_c}$ & 1 1 0 0 & 1/16 + 1/16 & $2$ \\
$\v{13}$ & $\sigma_1 e^{i 3/4 \sqrt{2} \varphi_c}$ & 0 1 1 0 & 1/16 + 9/16 & $2$ \\
$\v{14}$ & $\tau_1 e^{i 1/4 \sqrt{2} \varphi_c}$ & 0 0 1 1 & 9/16 + 1/16 & $2$ \\
$\v{15}$ & $\tau_1 e^{i 3/4 \sqrt{2} \varphi_c}$ & 1 0 0 1 & 9/16 + 9/16 & $2$ \\
\\
$\v{16}$ & $\sigma_2 e^{i 1/4 \sqrt{2} \varphi_c}$ &  1 1 0 0 & 1/16 + 1/16 & $2$ \\
$\v{17}$ & $\sigma_2 e^{i 3/4 \sqrt{2} \varphi_c}$ &  0 1 1 0  & 1/16 + 9/16 & $2$ \\
$\v{18}$ & $\tau_2 e^{i 1/4 \sqrt{2} \varphi_c}$ & 0 0 1 1 & 9/16 + 1/16 & $2$ \\
$\v{19}$ & $\tau_2 e^{i 3/4 \sqrt{2} \varphi_c}$ & 1 0 0 1 & 9/16 + 9/16 & $2$ \\
\\
$\v{20}$ & $\phi_2$ & 0 1 0 1 & $1/4 + 0$ & $2$ \\
$\v{21}$ & $\phi_2 e^{i 1/2 \sqrt{2} \varphi_c}$ & 1 0 1 0 & $1/4 + 1/4$ & $2$ \\
\hline
\end{tabular}
\caption{\label{Z2orbQP}
Quasiparticles obtained for $N = 4$ orbifold states by embedding the vertex
algebra into the $U(1)/Z_2 \times U(1)$ orbifold CFT. This vertex algebra contains
the electron operator, which is set to be $V_e = j e^{i \sqrt{2} \varphi_c}$, and
the operator $\cos(\sqrt{8} \varphi)$.  }
\end{table}

Now that we have two electron operators, we should be able to obtain 
a double-layer ideal wave function for these $N = 4$ states and 
an associated ideal Hamiltonian. We save an in-depth study of these
issues for future work. Based on the considerations
presented here, we expect that the vertex algebra generated by the two electron operators
has a unique, finite set of solutions for the quasiparticle structure constants, and therefore
that there is a corresponding gapped ideal Hamiltonian.

\subsubsection{$N = 5$}

The $\nu = 2/5$ orbifold FQH state, with $(N,q) = (5,0)$ is a fermionic state; 
a bosonic analog can be constructed at $\nu = 2/3$. In Tables \ref{N5b} and \ref{N5f}, 
we list the properties of these states. 

The electron operator is 
\begin{align}
\label{N5elOpOrb}
V_e = \phi_N^1 e^{i\sqrt{q+5/2} \phi_c},
\end{align}
where now $\phi^1_N$ has scaling dimension $h^{sc}_1 = 5/4$. The pattern of zeros
of this electron operator has also been studied in detail in \Ref{LW1024},
in the context of so-called $Z_2 | Z_4$ simple-current vertex algebra. We briefly
discuss this $N = 5$ case because it contains some novel features that did
not arise in the $N = 4$ analysis. In this $N  = 5$ case, we see that several
of the quasiparticle pattern of zeros solutions that are allowed by consistency 
conditions\cite{WW0809} do not appear (compare Table \ref{N5b} with Table VII of \Ref{LW1024})! 

This may be interpreted in the following way. The
single-layer ideal wave function with the pattern of zeros
of the operator in (\ref{N5elOpOrb}) is gapless, for the
same reason that the $N = 4$ case was gapless: the structure
constants of the vertex algebra for quasiparticles have a
continuous set of solutions. Driving the ideal wave function
away from the critical point, as discussed in the $N = 4$
example, corresponds to adding additional perturbations in
the ideal Hamiltonian and, from the perspective of the
vertex algebra, amounts to adding additional local operators
to the chiral algebra. The quasiparticles must all be local
with respect to this enlarged vertex algebra, however
certain quasiparticle pattern of zeros solutions may become
illegal as a result. Put another way, certain quasiparticle
pattern of zeros solutions that were allowed when the system
was thought of as a single-component system, may be illegal
if the chiral algebra is self-consistently enlarged in a
certain way to get a multi-component system. 

\begin{table}
\begin{center}
\begin{tabular}{clccc}
\hline
\multicolumn{5}{c}{Quasiparticles for $N = 5$ orbifold state ($\nu = 2/3$)} \\
\hline
 & & $\{n_{\gamma;l}\}$ & $h^{sc} + h^{ga}$ & quantum dim. \\
\hline
$\v 0$ & $\mathbb{I}$ & 2 0 2 0 0 0 & $0 + 0$ & 1\\
$\v 1$ & $e^{i2/3 \sqrt{\nu^{-1}} \phi_c}$ & 0 2 0 2 0 0 & $0 + 1/3$ & 1\\
$\v 2$ & $\phi_{N}^2 e^{i 1/3 \sqrt{\nu^{-1}} \phi_c}$ & 0 0 2 0 2 0 & $5/4 + 1/12$ & 1 \\
$\v 3$ & $j$ & 0 0 0 2 0 2 & $h = 1 + 0$ & 1 \\
$\v 4$ & $j e^{i2/3 \sqrt{\nu^{-1}} \phi_c}$ & 2 0 0 0 2 0 & $1 + 1/3$ & 1\\
$\v 5$ & $\phi_N^1 e^{i1/3 \sqrt{\nu^{-1}} \phi_c}$ & 0 2 0 0 0 2 & $5/4 + 1/12$ & 1\\
\\
$\v 6$ & $\phi_1  e^{i 1/3 \sqrt{\nu^{-1}} \phi_c}$ & 2 0 0 2 0 0 & $1/20 + 1/12$ & 2\\
$\v 7$ & $\phi_4$ & 0 2 0 0 2 0 & $4/5$ & 2 \\
$\v 8$ & $\phi_4 e^{i2/3 \sqrt{\nu^{-1}} \phi_c}$ & 0 0 2 0 0 2 & $4/5 + 1/3$ & 2 \\
\\
$\v 9$ & $\phi_2$ & 1 0 1 1 0 1 & $1/5 + 0$ & 2 \\
$\v{10}$ & $\phi_2 e^{i2/3 \sqrt{\nu^{-1}} \phi_c}$ & 1 1 0 1 1 0 & $1/5 + 1/3$ & 2 \\
$\v{11}$ & $\phi_3 e^{i 1/3 \sqrt{\nu^{-1}} \phi_c}$ & 0 1 1 0 1 1 & $9/20 + 1/12$ & 2 \\
\\
$\v{12}$ &  $\sigma_1 e^{i 1/2 \sqrt{\nu^{-1}}\phi_c}$ & 1 1 1 0 1 0 & $1/16 + 3/16$ & $\sqrt{5}$\\
$\v{13}$ &  $\sigma_2 e^{i 1/6 \sqrt{\nu^{-1}}\phi_c}$ & 0 1 1 1 0 1 & $1/16 + 1/48$ & $\sqrt{5}$\\
$\v{14}$ &  $\sigma_2 e^{i 5/6 \sqrt{\nu^{-1}}\phi_c}$ & 1 0 1 1 1 0 & $1/16 + 25/48$ & $\sqrt{5}$\\
$\v{15}$ &  $\tau_1 e^{i 1/2 \sqrt{\nu^{-1}}\phi_c}$ & 0 1 0 1 1 1& $9/16 + 3/16$ & $\sqrt{5}$\\
$\v{16}$ &  $\tau_2 e^{i 1/6 \sqrt{\nu^{-1}}\phi_c}$ & 1 0 1 0 1 1 & $9/16 + 1/48$ & $\sqrt{5}$\\
$\v{17}$ &  $\tau_2 e^{i 5/6 \sqrt{\nu^{-1}}\phi_c}$ & 1 1 0 1 0 1 & $9/16 + 25/48$ & $\sqrt{5}$\\
\hline
\end{tabular}
\end{center}
\caption{
\label{N5b}
Properties of the bosonic $(N,q) = (5,-1)$ orbifold states, at $\nu  = 2/3$.  Compare 
this to results from the $Z_2 | Z_4$ 
simple-current vertex algebra and pattern of zeros solutions studied in \Ref{LW1024}. 
$\phi_k = \cos(k/\sqrt{2N} \varphi)$.
}
\end{table}

\begin{table}
\begin{center}
\begin{tabular}{clccc}
\hline
\multicolumn{5}{c}{Quasiparticles for $N = 5$ orbifold state ($\nu = 2/5$)} \\
\hline
 & & $\{n_{\gamma;l}\}$ & $h^{sc} + h^{ga}$ & q. dim. \\
\hline
$\v 0$ & $\mathbb{I}$ &					 1  1  0  0  1  1  0  0  0  0 	& 0 + 0 & 1 \\
$\v 1$ & $e^{i 2/5 \sqrt{\nu^{-1}}\phi_c}$ &	 0  1  1  0  0  1  1  0  0  0 & 0 + 1/5 & 1\\
$\v 2$ & $e^{i 4/5 \sqrt{\nu^{-1}}\phi_c}$ &		 0  0  1  1  0  0  1  1  0  0 & 0 + 4/5 & 1\\
$\v 3$ & $\phi_N^2 e^{i 1/5 \sqrt{\nu^{-1}}\phi_c}$ &		 0  0  0  1  1  0  0  1  1  0 & 5/4 + 1/20 & 1 \\
$\v 4$ & $\phi_N^2 e^{i 3/5 \sqrt{\nu^{-1}}\phi_c}$ &		 0  0  0  0  1  1  0  0  1  1 & 5/4 + 9/20 & 1\\
$\v 5$ & $j$ &					 1  0  0  0  0  1  1  0  0  1 & 1 + 0 & 1\\
$\v 6$ & $j e^{i 2/5 \sqrt{\nu^{-1}}\phi_c}$ &		 1  1  0  0  0  0  1  1  0  0 & 1 + 1/5 & 1\\
$\v 7$ & $j e^{i 4/5 \sqrt{\nu^{-1}}\phi_c}$ &		 0  1  1  0  0  0  0  1  1  0 & 1 + 4/5 & 1\\
$\v 8$ & $\phi_N^1 e^{i 1/5 \sqrt{\nu^{-1}}\phi_c}$ &		 0  0  1  1  0  0  0  0  1  1 & 5/4 + 1/20 & 1\\
$\v 9$ & $\phi_N^1 e^{i 3/5 \sqrt{\nu^{-1}}\phi_c}$ &		 1  0  0  1  1  0  0  0  0  1 & 5/4 + 9/20 & 1\\
\\
$\v{10}$ & $\phi_1 e^{i 1/5 \sqrt{\nu^{-1}}\phi_c}$	 &	 1  1  0  0  0  1  1  0  0  0 & 1/20 + 1/20 & 2\\
$\v{11}$ & $\phi_1 e^{i 3/5 \sqrt{\nu^{-1}}\phi_c}$	 & 0  1  1  0  0  0  1  1  0  0 & 1/20 + 9/20 & 2\\
$\v{12}$ & $\phi_4$ &		 0  0  1  1  0  0  0  1  1  0 & 4/5 + 0 & 2\\
$\v{13}$ & $\phi_4 e^{i 2/5 \sqrt{\nu^{-1}}\phi_c}$	&  0  0  0  1  1  0  0  0  1  1 & 4/5 + 1/5 & 2\\
$\v{14}$ & $\phi_4 e^{i 4/5 \sqrt{\nu^{-1}}\phi_c}$	&	 1  0  0  0  1  1  0  0  0  1 & 4/5 + 4/5 & 2\\
\\
$\v{15}$ & $\phi_2$ &					 0  1  0  0  1  0  1  0  0  1 & 1/5 + 0 & 2\\
$\v{16}$ & $\phi_2 e^{i 2/5 \sqrt{\nu^{-1}}\phi_c}$	&	 1  0  1  0  0  1  0  1  0  0 & 1/5 + 1/5 & 2\\
$\v{17}$ & $\phi_2 e^{i 4/5 \sqrt{\nu^{-1}}\phi_c}$	&	 0  1  0  1  0  0  1  0  1  0 & 1/5 + 4/5 & 2\\
$\v{18}$ & $\phi_3 e^{i 1/5 \sqrt{\nu^{-1}}\phi_c}$	&	 0  0  1  0  1  0  0  1  0  1 & 9/20 + 9/20 & 2\\
$\v{19}$ & $\phi_3 e^{i 3/5 \sqrt{\nu^{-1}}\phi_c}$	&	 1  0  0  1  0  1  0  0  1  0 & 9/20 + 9/20 & 2\\
\\
$\v{20}$ & $\sigma_1 e^{i 3/10 \sqrt{\nu^{-1}}\phi_c}$ &		 1  0  1  0  1  0  0  1  0  0 & 1/16 + 0.113 & $\sqrt{5}$ \\
$\v{21}$ & $\sigma_1 e^{i 7/10 \sqrt{\nu^{-1}}\phi_c}$ &		 0  1  0  1  0  1  0  0  1  0 & 1/16 + 0.613 & $\sqrt{5}$ \\
$\v{22}$ & $\sigma_2 e^{i 1/10 \sqrt{\nu^{-1}}\phi_c}$ &		 0  0  1  0  1  0  1  0  0  1 & 1/16 + 0.012 & $\sqrt{5}$ \\
$\v{23}$ & $\sigma_2 e^{i 5/10 \sqrt{\nu^{-1}}\phi_c}$ &		 1  0  0  1  0  1  0  1  0  0 & 1/16 + 0.312 & $\sqrt{5}$ \\
$\v{24}$ & $\sigma_2 e^{i 9/10 \sqrt{\nu^{-1}}\phi_c}$ &		 0  1  0  0  1  0  1  0  1  0 & 1/16 + 1.012 & $\sqrt{5}$ \\
$\v{25}$ & $\tau_1 e^{i 3/10 \sqrt{\nu^{-1}}\phi_c}$ &		 0  0  1  0  0  1  0  1  0  1 & 9/16 + 0.113 & $\sqrt{5}$ \\
$\v{26}$ & $\tau_1 e^{i 7/10 \sqrt{\nu^{-1}}\phi_c}$	&	 1  0  0  1  0  0  1  0  1  0 & 9/16 + 0.613 & $\sqrt{5}$ \\
$\v{27}$ & $\tau_2 e^{i 1/10 \sqrt{\nu^{-1}}\phi_c}$	&	 0  1  0  0  1  0  0  1  0  1 & 9/16 + 0.012 & $\sqrt{5}$ \\
$\v{28}$ & $\tau_2 e^{i 5/10 \sqrt{\nu^{-1}}\phi_c}$	&	 1  0  1  0  0  1  0  0  1  0 & 9/16 + 0.312 & $\sqrt{5}$ \\
$\v{29}$ & $\tau_2 e^{i 9/10 \sqrt{\nu^{-1}}\phi_c}$	&	 0  1  0  1  0  0  1  0  0  1 & 9/16 + 1.012 & $\sqrt{5}$ \\
\hline
\end{tabular}
\end{center}
\caption{
\label{N5f}
Properties of the fermionic $(N,q) = (5,0)$ orbifold states, at $\nu  = 2/5$.
$\phi_k = \cos(k/ \sqrt{2N} \varphi)$.
}
\end{table}

\subsubsection*{Conclusion}

This concludes our analysis of the orbifold FQH states from the point of
view of ideal wave functions and the vertex algebra/pattern of zeros
approach. The orbifold FQH states provide the first concrete examples 
in which the operators of the edge CFT have ``sick'' pattern of zeros solutions. 
As a result, we find that these states yield profound
insights into the vertex algebra framework. Namely, when a
certain pattern of zeros solution appears to describe a
gapless state (due to a continuum of solutions to the
quasiparticle structure constants in the vertex algebra),
this means that generically there may be a way to
self-consistently enlarge the vertex algebra, which
physically corresponds to condensing new operators and
driving the ideal Hamiltonian away from a critical point.
Then the newly enlarged vertex algebra may have a finite
number of quasiparticle solutions and there may be a
multilayer ideal wave function that captures the topological
order of the resulting states. Thus all of the pattern of
zeros solutions, even when they naively appear to be
describing gapless phases, are ultimately relevant in
describing incompressible FQH states! 

An important direction now is to put the above ideas on a
more concrete footing in the vertex algebra framework in
order to, for instance, derive the incompressible ideal wave
functions that do capture the topological order of the
orbifold FQH states. 

\section{Relation to experiments and relevance to $\nu = 8/3$ and $12/5$}
\label{expSec}

In both double-layer and wide single-layer quantum wells, several of the $(ppq)$ 
states, such as the $(331)$ and $(330)$ states, have been routinely realized 
experimentally.\cite{LJS9792,SM9405} The study presented here suggests that by varying material parameters
such as the interlayer tunneling/repulsion, it could be possible
to tune through a continuous quantum phase transition into a non-Abelian FQH state.

Since the transition is driven by the condensation of an electrically neutral 
boson, the charge gap remains nonzero through the transition, which would make
detection of the transition difficult through charge transport experiments. 
Some possible experimental probes are as follows. 

The most obvious physical consequence of this transition is that the bulk should
become a thermal conductor at the transition, because while the charge gap remains, a 
neutral mode becomes gapless at the critical point. This would also have a pronounced
effect on edge physics; near the transition, the velocity of a neutral mode 
approaches zero, until at the transition it becomes a gapless excitation in the bulk. 
Thus this transition should also be detectable through edge tunneling experiments. 
Furthermore, as discussed in Section \ref{anyonCond}, the transition should be accompanied
by interlayer density fluctuations. Since the density fluctuations carry an electric dipole moment,
they can in principle be observed through surface acoustic phonons.\cite{WR9344} 

One useful physical distinction between the bilayer Abelian states and the orbifold
non-Abelian states are that when $N = p-q$ is odd, the minimal electric charge of the
quasiparticles becomes halved in the orbifold phase. Thus, for example the quasiparticle
minimal charge can be measured as the interlayer tunneling and interlayer thickness
are tuned in a two-component $(330)$ state. An observation of a change in the minimal
quasiparticle electric charge from $e/3$ to $e/6$ would indicate a transition to the
non-Abelian phase.  

Another implication of the results here applies to the single-layer plateaus
that have been observed at $\nu = 8/3$ and $\nu = 12/5$.\cite{XP0409} Currently, it is
believed that the FQH plateaus seen in single layer samples
at $\nu = 8/3 = 2 + 2/3$ and $\nu = 12/5 = 2 + 2/5$ might 
be exotic non-Abelian states.\cite{CK0801} There are a number of 
candidate states, including the particle-hole conjugate of the $Z_3$ 
parafermion (Read-Rezayi) state\cite{RR0906} and some hierarchy states formed over the 
Pfaffian state.\cite{BS0823, BF0964}

Our study suggests another set of possible states. The orbifold states presented
here are neighbors in the phase diagram to more conventional states, such as 
the $(330)$ and $(550)$ states. These states can exist at $\nu = 8/3$ and 
$12/5$, respectively -- in fact, experiments on wide single layer quantum 
wells have seen plateaus at $\nu = 8/3$. The fact that the orbifold FQH states
are neighbors in the phase diagram to these more conventional bilayer states
means that in single-layer samples, the orbifold FQH states should be considered
as possible candidates to explain the observed plateaus. 

\section{Summary, conclusions, and outlook}
\label{conclusion}

\begin{figure}
\includegraphics[scale=0.46]{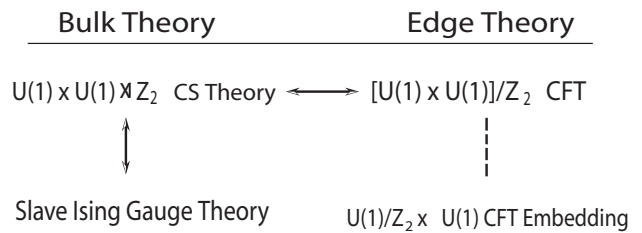}
\caption{
\label{orbSchematic}
Here we try to illustrate the different ways of describing the topological order of
the orbifold FQH states and how they are related. The $U(1) \times U(1) \rtimes Z_2$
CS theory is a bulk topological field theory, but we do not know how to compute all
of the topological properties from this theory. Closely related is the slave Ising
theory gauge theory presented in Section \ref{slaveSect}, which we believe provides a lattice
regularization of the $U(1) \times U(1) \rtimes Z_2$ CS theory. The spectrum of the 
edge theory for these bulk topological theories is conjectured to be generated by the
electron operator $\Psi_e = \cos (\sqrt{N/2} \varphi) e^{i \sqrt{(p+q)/2} \phi_c}$. We
also conjecture that this edge theory is equivalent to what one would obtain 
for the edge theory by taking the electron operator to be 
$\Psi_e = \phi_N^1 e^{i \sqrt{(p+q)/2} \phi_c}$ and embedding it into the holomorphic 
half of the $Z_2$ orbifold $\times U(1)_{charge}$ CFT. 
}
\end{figure}

\begin{table*}
\begin{center}
\begin{tabular}{llccccc}
\hline
                   \;\;\;\;\;                  &    & No. of QP & Quantum dimensions & Charges & Fusion rules & Scaling dimensions \\
\hline
\multirow{2}{*}{Bulk} &  $U(1) \times U(1) \rtimes Z_2$ & $\checkmark$          & $\checkmark$       & Some             & Some         & Some  \\
 & Slave Ising theory              &                       &                    & $\checkmark$     &              &      \\
\\
\\
\multirow{2}{*}{Edge} &   $U(1)/Z_2 \times U(1)$ CFT prescription & $\checkmark$          & $\checkmark$       & $\checkmark$     & $\checkmark$ & $\checkmark$ \\
& $[U(1) \times U(1)]/Z_2$ CFT      &     $\checkmark$                  &                    &                  &              &   Some         \\
\hline
\end{tabular}
\caption{
\label{orbSuc}
Summary of the successes of various descriptions of the orbifold FQH states. 
}
\end{center}
\end{table*}

In this paper, we have introduced a set of FQH phases, dubbed the orbifold FQH states,
and studied phase transitions between them and conventional Abelian bilayer phases.
The orbifold states are labelled by two parameters $(N,q)$ and exist at filling fraction
$\nu = 2/(N + 2q)$. The bulk low energy effective field theory for these phases is
the $U(1) \times U(1) \rtimes Z_2$ CS theory. Their edge CFT is a $[U(1) \times U(1)]/Z_2$
orbifold CFT with central charge $c = 2$. These orbifold phases contain an electrically
neutral boson whose condensation drives a continuous quantum phase transition to
the bilayer $(ppq)$ states. In the $U(1) \times U(1) \rtimes Z_2$ CS theory, this neutral
boson carries $Z_2$ gauge charge and so the effective theory near the transition -- where
the neutral boson has low energy compared to all other excitations -- is 
a $Z_2$ gauged Ginzburg-Landau theory, which implies that the transition is in the 3D Ising
universality class. 

We introduced a slave-particle gauge theory formulation of these states, which shows how
to interpolate between the Abelian bilayer states and the orbifold states. This description
provides an interesting example in which $Z_2$ fractionalization leads to
non-Abelian topological phases. Finally, we saw that the existence of these states sheds
considerable light on the pattern of zeros/vertex algebra framework for characterizing
ideal FQH wave functions. The orbifold states provide the first examples in which the
sick pattern of zeros solutions are actually relevant for describing incompressible FQH states. 

The calculation of the full topological quantum numbers of the quasiparticles relies on 
a prescription in which we embed  the electron operator in the $U(1)/Z_2 \times U(1)$ 
CFT. We have not proven rigorously that the results are equivalent to the $[U(1) \times U(1)]/Z_2$
CFT. Let us briefly summarize the successes 
of the various descriptions of the orbifold FQH states, as shown in Table \ref{orbSuc}.
The bulk $U(1) \times U(1) \rtimes Z_2$ CS theory can be used to compute to the number of quasiparticles
and the quantum dimensions of all of the quasiparticles, which can yield the ground state degeneracy
on genus $g$ surfaces. Based on the relation to the neighboring $(ppq)$ states, we can deduce the
charges and twists/scaling dimensions of the quasiparticles that have quantum dimension 
1 and $2$, but not those of the $Z_2$ vortices. This relation to the $(ppq)$ states also allows us to deduce 
certain properties of the fusion rules. Furthermore, by studying the $Z_2$ vortices in 
detail, we can deduce some information about their fusion rules as well from the $U(1) \times U(1) \rtimes Z_2$ CS theory. 
The bulk $U(1) \times U(1) \rtimes Z_2$ CS theory is closely related to the slave Ising theory
introduced in Section \ref{slaveSect}, which allows us to compute the charges of the $Z_2$ vortices. 
However from these bulk theories we do not know how to compute the twists of the $Z_2$ vortices or
all of the fusion rules of the quasiparticles.

The edge theory of the orbifold states is the $[U(1) \times U(1)]/Z_2$ CFT, and the electron
operator is $\Psi_e =  \cos (\sqrt{N/2} \varphi_-) e^{i \sqrt{(p+q)/2} \phi_+}$. However, using
this operator we currently can only compute the pattern of zeros of the electron operator, which
yields the scaling dimensions of the Abelian quasiparticles in the theory.\cite{WW0809, BW0932} 
Based on a close relation to the $Z_2$ orbifold chiral algebra, we conjecture that the 
topological order can be completely described by setting the electron operator to be
$\phi_N^1 e^{i \sqrt{\nu^{-1}} \phi_c}$ and embedding the electron operator into the 
$U(1)_{2N}/Z_2 \times U(1)$ CFT. From this prescription, we can compute all
topological properties, and they agree with all quantities that can be 
computed in any other ways.  

There are many directions for future research. Conceptually, perhaps the most
interesting and important would be to further the understanding of these states through the perspective
of ideal wave functions. The ideas presented in this paper regarding the pattern
of zeros and vertex algebra approaches to ideal wave functions are preliminary and need
to be borne out by more concrete calculations. These orbifold states are currently the 
only FQH states for which we do not have an ideal wave function with the same 
topological properties. 

Another direction is to fill in the logical leaps in the analysis presented here, such as deriving
all properties of the edge theory directly from the $[U(1) \times U(1)]/Z_2$
orbifold CFT, without using the prescription in terms of the 
$U(1)_{2N}/Z_2 \times U(1)$ CFT. Similarly, ways could be developed for computing
more topological properties directly from the $U(1) \times U(1) \rtimes Z_2$ CS theory.

An important direction in order to make contact with experiments is to use 
the projected trial wave functions developed here using the slave-particle 
gauge theory in order to numerically analyze where in
the bilayer phase diagram these orbifold FQH states may be favorable. Given their
close proximity in the phase diagram to conventional Abelian states that are routinely
realized in experimental samples, it is possible that experiments might actually probe
this transition. Furthermore, their proximity in the phase diagram to conventional states
warrants their inclusion as possible candidates for explaining the single-layer 
plateaus observed experimentally at $\nu = 8/3$ and $12/5$ and whose ultimate nature 
remains mysterious. 

Finally, we note that while the mathematical theory of boson condensation in
topological phases has been recently understood, our study provides a physical
context and understanding of the physical properties of such phase transitions
in the simplest case of the condensed boson having $Z_2$ fusion rules. More 
generally, it is known that in tensor category theory, one can start with a
tensor category $\mathcal{C}_0$ and mod out by a subcategory $\mathcal{C}$
where $\mathcal{C}$ contains the fusion subalgebra that is generated by a 
boson in the theory. \Ref{BS0916} studies many general properties of the topological quantum
numbers of such a transition. One direction is to extend our work by studying
the physics of such transitions: the physical contexts in which they occur, 
effective field theories that describe the two phases, and the nature
of the transitions. 

As a concluding remark we would like to mention that some of the mathematics of the topological
order of these orbifold FQH states was recently studied in \Ref{RW1041}.

We thank Michael Levin, Zhenghan Wang, and Brian Swingle for discussions. We also thank
Tsinghua University for hospitality while part of this work was completed. This research was supported by
NSF Grant No.DMR-1005541. MB is currently supported by a fellowship from the Simons Foundation. 

\appendix
\section{$U(1)/Z_2$ orbifold CFT }
\label{orbAppendix}

Since the $U(1)/Z_2$ orbifold at $c = 1$ plays an important role in
understanding the topological properties of the orbifold FQH states, 
here we will give a brief account of some of its properties. 
The information here is taken from \Ref{DV8985},
where a more complete discussion can be found. 

The $U(1)/Z_2$ orbifold CFT, at central charge $c = 1$, is the theory 
of a scalar boson $\varphi$, compactified at a radius $R$, so that 
$\varphi \sim \varphi + 2\pi R$, and with an additional $Z_2$ gauge 
symmetry: $\varphi \sim -\varphi$. When $\frac{1}{2} R^2$ is 
rational, \it i.e. \rm $\frac{1}{2} R^2 = p/p'$, 
with $p$ and $p'$ coprime, then it is useful to
consider an algebra generated by the fields $j = i \partial \varphi$,
and $e^{\pm i \sqrt{2N} \varphi}$, for $N = p p'$. This algebra is referred to
as an extended chiral algebra. The infinite number of Virasoro primary fields
in the $U(1)$ CFT can now be organized into a finite number of representations 
of this extended algebra $\mathcal{A}_N$. There are $2N$ of these 
representations, and the primary fields are written as $V_k = e^{ik\varphi/\sqrt{2N}}$, 
with $k = 0, 1, \cdots, 2N-1$. The $Z_2$ action takes $V_k \rightarrow V_{2N-k}$.

In the $Z_2$ orbifold, one now considers representations of the smaller
algebra $\mathcal{A}_N/Z_2$. This includes the $Z_2$ invariant combinations of the
original primary fields, which are of the form $\phi_k = \cos(k \varphi/\sqrt{2N})$;
there are $N+1$ of these. In addition, there are 6 new primary fields. The gauging
of the $Z_2$ allows for twist operators that are not local with respect to the fields
in the algebra $\mathcal{A}_N/Z_2$, but rather local up to an element of $Z_2$. It turns
out that there are two of these twisted sectors, and each sector contains one field
that lies in the trivial representation  of the $Z_2$, and one field that lies
in the non-trivial representation of $Z_2$. These twist fields are labelled
$\sigma_1$, $\tau_1$, $\sigma_2$, and $\tau_2$. In addition to these, an in-depth
analysis \citep{DV8985} shows that the fixed points of the $Z_2$ action in the original $U(1)$ 
theory split into a $Z_2$ invariant and a non-invariant field. We have already 
counted the invariant ones in our $N+1$ invariant fields, which leaves 2 new
fields. One fixed point is the identity sector, corresponding to $V_0$, which 
splits into two sectors: $1$, and $j = i \partial \varphi$. The other
fixed point corresponds to $V_N$. This splits into two primary fields, which 
are labelled as $\phi_N^i$ for $i = 1,2$ and which have scaling dimension $N/4$.
In total, there are $N+7$ primary fields in the $Z_2$ rational orbifold at ``level'' $2N$.
These fields and their properties are summarized in Table \ref{Z2orbFields}.
\begin{table}
\begin{center}
\begin{tabular}{ccc}
\hline
Label & Scaling Dimension & Quantum Dimension \\
\hline
$\mathbb{I}$ & 0 & 1 \\
$j$ & 1 & 1 \\
$\phi_N^1$ & $N/4$ & 1 \\
$\phi_N^2$ & $N/4$ & 1 \\
\\
$\sigma_1$ & 1/16 & $\sqrt{N}$ \\
$\sigma_2$ & 1/16 & $\sqrt{N}$ \\
$\tau_1$ & 9/16 & $\sqrt{N}$ \\
$\tau_2$ & 9/16 & $\sqrt{N}$ \\
\\
$\phi_k$ & $k^2/4N$ & $2$ \\
\hline
\end{tabular}
\end{center}
\caption{\label{Z2orbFields}
Primary fields in the $U(1)_{2N}/Z_2$ orbifold CFT. The label $k$ runs
from $1$ to $N-1$.}
\end{table}

This spectrum for the $Z_2$ orbifold is obtained by first computing the
partition function of the full $Z_2$ orbifold CFT defined on a torus, 
including both holomorphic and anti-holomorphic parts. Then, the partition
function is decomposed into holomorphic blocks, which are conjectured to
be the generalized characters of the $\mathcal{A}_N/Z_2$ chiral algebra. 
This leads to the spectrum listed in Table \ref{Z2orbFields}. 
The fusion rules and scaling dimensions for these primary fields are obtained 
by studying the modular transformation properties of the characters.

The fusion rules are as follows. For $N$ even:
\begin{align}
j \times j  &=1,
\nonumber \\
\phi^i_N \times \phi^i_N &= 1,
\nonumber \\
\phi^1_N \times \phi_N^2 &= j.
\end{align}
As mentioned in \Ref{DV8985}, the vertex operators $\phi_k$ have a fusion algebra
consistent with their interpretation as $\cos \frac{k}{\sqrt{2N}} \varphi$,
\begin{align}
\phi_k \times \phi_{k'} &= \phi_{k+k'} + \phi_{k - k'} \;\;\;\;\; (k' \neq k, N- k),
\nonumber \\
\phi_k \times \phi_k &= 1 + j + \phi_{2k},
\nonumber \\
\phi_{N-k} \times \phi_k &= \phi_{2k} + \phi^1_N + \phi^2_N,
\nonumber \\
j \times \phi_k &= \phi_k.
\end{align}
\begin{align}
\sigma_i \times \sigma_i &= 1 + \phi_N^i + \sum_{k \text{ even}} \phi_k,
\nonumber \\
\sigma_1 \times \sigma_2 &= \sum_{k \text{ odd}} \phi_k,
\nonumber \\
j \times \sigma_i &= \tau_i
\end{align}

\begin{table}[t]
\begin{tabular}{ccc}
\hline
$Z_2$ Orb. field & Scaling Dimension, $h$ & $Z_4$ parafermion field \\
\hline
\hline
$1$ & $0$ & $\Phi^0_0$ \\
$j$ & $1$ & $\Phi^0_4$ \\
$\phi^1_N$ & $3/4$ & $\Phi^0_2$ \\
$\phi^2_N$ & $3/4$ & $\Phi^0_6$ \\
$\phi_1$ & $1/12$ & $\Phi^2_2$ \\
$\phi_2$ & $1/3$ & $\Phi^2_0$ \\
$\sigma_1$ & $1/16$ & $\Phi^1_1$ \\
$\sigma_2$ & $1/16$ & $\Phi^1_{-1}$ \\
$\tau_1$ & $9/16$ & $\Phi^1_3$ \\
$\tau_2$ & $9/16$ & $\Phi^1_5$ \\
\hline
\end{tabular}
\caption{
\label{Z2fieldsN3}
Primary fields in the $Z_2$ orbifold for $N=3$, their 
scaling dimensions, and the $Z_4$ parafermion fields that they correspond to.}
\end{table}

For $N$ odd, the fusion algebra of 1, $j$, and $\phi_N^i$ is
$Z_4$:
\begin{align}
j \times j &= 1,
\nonumber \\
\phi_N^1 \times \phi_N^2 &= 1,
\nonumber \\
\phi_N^i \times \phi_N^i &= j.
\end{align}
The fusion rules for the twist fields become:
\begin{align}
\sigma_i \times \sigma_i &= \phi_N^i + \sum_{k \text{ odd}} \phi_k,
\nonumber \\
\sigma_1 \times \sigma_2 &= 1 + \sum_{k \text{ even}} \phi_k.
\end{align}
The fusion rules for the operators $\phi_k$ are unchanged. 

For $N = 1$, it was observed that the $Z_2$ orbifold is equivalent
to the $U(1)_8$ Gaussian theory. For $N = 2$, it was observed 
that the $Z_2$ orbifold is equivalent to two copies of the 
Ising CFT. For $N = 3$, it was observed that the $Z_2$ orbifold 
is equivalent to the $Z_4$ parafermion CFT of Zamolodchikov
and Fateev.\citep{ZF8515}

In Tables \ref{Z2fieldsN2} and \ref{Z2fieldsN3} we list the
fields from the $Z_2$ orbifold for $N=2$ and $N=3$, their
scaling dimensions, and the fields in the $Ising^2$ or $Z_4$ 
parafermion CFTs that they correspond to.

\begin{table}
\begin{tabular}{ccc}
\hline
$Z_2$ Orb. field & Scaling Dimension, $h$ & $Ising^2$ fields \\
\hline
\hline
$1$ & $0$ & $\mathbb{I} \otimes \mathbb{I}$ \\
$j$ & $1$ & $\psi \otimes \psi$ \\
$\phi^1_N$ & $1/2$ & $\mathbb{I} \otimes \psi$ \\
$\phi^2_N$ & $1/2$ & $\psi \otimes \mathbb{I}$ \\
$\phi_1$ & $1/8$ & $\sigma \otimes \sigma$ \\
$\sigma_1$ & $1/16$ & $\sigma \times \mathbb{I}$ \\
$\sigma_2$ & $1/16$ & $\mathbb{I} \otimes \sigma$ \\
$\tau_1$ & $9/16$ & $\sigma \otimes \psi$ \\
$\tau_2$ & $9/16$ & $\psi \otimes \sigma$ \\
\hline
\end{tabular}
\caption{
\label{Z2fieldsN2}
Primary fields in the $Z_2$ orbifold for $N=2$, their 
scaling dimensions, and the fields from Ising$^2$ to which they
correspond.}
\end{table}

\section{$Z_N$ transitions between Abelian states}
\label{ZNtrans}

Our analysis of the $N = 1$ orbifold states in Section \ref{N1Orb}
revealed a series of Abelian FQH states that can apparently
undergo $Z_m$ phase transitions to other Abelian FQH states. 

In particular, consider the following two-component states 
with $K$-matrix and charge vector $\v q$ given by
\begin{align}
K = \left( \begin{matrix}
2m^2 & m \\
m & q+1 \\
\end{matrix} \right)
\;\;\;\;\;
\v q = \left( \begin{matrix}
0 \\
1 \\
\end{matrix} \right)
\end{align}
For $m > 1$, these states have a neutral boson $\phi$ with the fusion rule
\begin{align}
\phi^m = \mathbb{I}.
\end{align}
To see this, observe that $\phi$ can be described by the integer vector $l_\phi^T = (2m, 1)$. From the
formula $Q_\phi = \v q^T K^{-1} l_\phi = 0$ we find that $\phi$ is electrically neutral,
while from $\theta_\phi / \pi =l_\phi^T K^{-1} l_\phi = \text{ even}$ we find that
$\phi$ is a boson. Finally, from the fact that $m l_\phi^T = (2m^2, m)$, which is the
first row of the $K$-matrix, we find that $\phi^m$ is a local excitation, \it ie \rm
$\phi^m = \mathbb{I}$. 
 
Based on the analysis in Section \ref{N1Orb},
we expect that the condensation of $\phi$ will yield the $m = 1$ states
and that the transition is in the $Z_m$ universality class. In the case
$m = 2$, these are the $N = 1$ orbifold FQH states, which have non-Abelian
analogs for more general $N$. Also, in the case $m = 2$ there is a $U(1) \times U(1) \rtimes Z_2$
CS description that makes the appearance of this discrete $Z_2$ structure
explicit. 

We currently do not know whether for $m > 2$ there are also 
non-Abelian analogs that are separated from a bilayer Abelian 
phase by a $Z_m$ transition. We also do not know whether
there is a way to describe these states in terms of a CS theory
with a gauge group that makes the $Z_m$ structure explicit,
as there is for $m  = 2$.

\section{Slave rotor construction}
\label{slaveRotorApp}

While the slave Ising construction presented above is sufficient to describe the 
bilayer Abelian $(ppq)$ states and the non-Abelian orbifold FQH states, it is 
a ``minimal'' slave-particle gauge theory in the sense that it only captures the
minimal amount of fluctuations about a given mean-field state in order to see
the possibility of the two phases. It is possible to improve the slave-particle
description by including more of these fluctuations about the mean-field states
and probing a larger part of the Hilbert space. This can be done by promoting
the above slave Ising theory to the following slave rotor description. 

We rewrite the electron operators in the following way:
\begin{align}
\Psi_{i+} &= c_{i+},
\nonumber\\
\Psi_{i-} &= e^{i\phi_i} c_{i-}.
\end{align}
In such a construction, we have a $U(1)$ gauge symmetry 
associated with the following local transformations:
\begin{align}
\phi_i \rightarrow \phi_i + \alpha, \;\;\;\;\; c_{i-} \rightarrow e^{-i \alpha} c_{i-}.
\end{align}
This means that the physical states must satisfy
\begin{align}
\label{rotorCon1}
e^{i \alpha \hat{L}_i - i \alpha n_{c_{i-}}} = 1, 
\end{align}
for any $\alpha$ (there is an arbitrary $U(1)$ phase factor that we
have set to unity here). The angular momentum $\hat{L_i} \propto i \partial \phi_i$ 
is conjugate to the field $\phi_i$. (\ref{rotorCon1}) implies
\begin{align}
\label{rotorCon2}
\hat{L}_i - n_{c_{i-}}= 0
\end{align} 

Note that we will actually want to do a further slave-particle decomposition
into partons, as in (\ref{slaveIsing}). For example, for $q = 0$, we decompose $c_{\pm}$ as
\begin{align}
c_{\pm i} = \prod_{a=1}^N \psi_{ai} \pm \prod_{a=N+1}^{2N} \psi_{ai} .
\end{align}
In this case, the gauge symmetry associated with translating $\phi_i$ is actually only
a $Z_2$ symmetry:
\begin{align}
\phi_i \rightarrow \phi_i + \pi, \;\;\;\;\; \psi_i \leftrightarrow \psi_{N+i}.
\end{align}
In this case, the constraint on the rotor is actually
$\hat{L}_i - n_{c_{i-}} = $ even. Or, alternatively:
\begin{align}
(-1)^{\hat{L}_i + n_{c_{i-}}} = 1.
\end{align}

Let us set $b_i \equiv e^{i \phi_i}$. Substituting into a model Hamiltonian that includes
hopping between sites in the same layer, between sites in different layers, and various
interaction terms, we obtain:
\begin{widetext}
\begin{align}
H_{kin} + H_{tun} =& \sum_{ij} (t_{ij} + t_{ji}^* +T_{ij} +T_{ji}^*) c_{i+}^{\dagger} c_{j+} 
+ \sum_{ij} (t_{ij} + t_{ji}^* - T_{ij} - T_{ji}^*) b_i^* b_j c_{i-}^{\dagger} c_{j-})
\nonumber \\
H_{int} =& \frac{1}{2}\sum_{ij} (U_{ij} + V_{ij}) 
: (n_{i+} n_{j+} + n_{i+} n_{j-}  + n_{i-} n_{j+} + n_{i-} n_{j-})  :
\nonumber \\
& + \frac{1}{2} \sum_{ij} (U_{ij} - V_{ij})  (b_j b_i c_{+i}^{\dagger}  c_{+j}^{\dagger} c_{-j} c_{-i} 
+ b_i^* b_j c_{-i}^{\dagger}  c_{+j}^{\dagger} c_{-j} c_{+i}
+ b_i^* b_j^* c_{-i}^{\dagger}  c_{-j}^{\dagger} c_{+ j} c_{+i} +  b_i b_j^* c_{+i}^{\dagger}  c_{-j}^{\dagger} c_{+j} c_{-i}) 
\end{align}
\end{widetext}
Note that the above Hamiltonian does not preserve a global $U(1)$ symmetry associated
with arbitrary translations of $\phi_i$; there is only a $Z_2$ symmetry. 
Thus there are only two distinct phases.
The first one is smoothly connected to a situation in which
\begin{align}
\langle e^{i \phi} \rangle \neq 0 ,
\end{align}
and the second one is smoothly connected to a situation in which 
\begin{align}
\langle e^{i 2 \phi} \rangle \neq 0 .
\end{align}
The first possibility breaks the $Z_2$ gauge symmetry, while the second one preserves it. 
These two possibilities describe precisely the same two phases as the slave
Ising theory described above. In the first case, suppose we set
$ e^{i\phi} = 1$. Then, we are left with the parton construction for the $(ppq)$
states. In the $Z_2$ unbroken phase, we may set $e^{i 2 \phi} = 1$, 
so that $e^{i \phi} = \pm 1 \equiv s^z_i$. Thus, these two phases that we can access
in the slave rotor approach are the same phases that we can access from the slave
Ising approach. The $Z_2$ broken phase corresponds to the bilayer $(ppq)$ states,
while the $Z_2$ unbroken phase corresponds to the orbifold FQH states. 
The slave rotor approach has the advantage of probing more fluctuations around
the mean-field states because more of the Hilbert space is being accessed in 
this decomposition. This may allow for more reliable calculations of the phase
diagram. 

As in the slave Ising construction, this slave rotor construction also provides
trial projected wave functions, but provides a larger space of possible
trial wave functions that capture the behavior of each of the two phases.


\begin{thebibliography}{53}
\expandafter\ifx\csname natexlab\endcsname\relax\def\natexlab#1{#1}\fi
\expandafter\ifx\csname bibnamefont\endcsname\relax
  \def\bibnamefont#1{#1}\fi
\expandafter\ifx\csname bibfnamefont\endcsname\relax
  \def\bibfnamefont#1{#1}\fi
\expandafter\ifx\csname citenamefont\endcsname\relax
  \def\citenamefont#1{#1}\fi
\expandafter\ifx\csname url\endcsname\relax
  \def\url#1{\texttt{#1}}\fi
\expandafter\ifx\csname urlprefix\endcsname\relax\def\urlprefix{URL }\fi
\providecommand{\bibinfo}[2]{#2}
\providecommand{\eprint}[2][]{\url{#2}}

\bibitem[{\citenamefont{Wen}(2004)}]{Wen04}
\bibinfo{author}{\bibfnamefont{X.-G.} \bibnamefont{Wen}},
  \emph{\bibinfo{title}{Quantum Field Theory of Many-Body Systems -- From the
  Origin of Sound to an Origin of Light and Electrons}}
  (\bibinfo{publisher}{Oxford Univ. Press}, \bibinfo{address}{Oxford},
  \bibinfo{year}{2004}).

\bibitem[{\citenamefont{Wang}(2010)}]{W10}
\bibinfo{author}{\bibfnamefont{Z.}~\bibnamefont{Wang}},
  \emph{\bibinfo{title}{Topological Quantum Computation}}
  (\bibinfo{publisher}{American Mathematical Society}, \bibinfo{year}{2010}).

\bibitem[{\citenamefont{Preskill}(2004)}]{P04}
\bibinfo{author}{\bibfnamefont{J.}~\bibnamefont{Preskill}}
  (\bibinfo{year}{2004}),
  \urlprefix\url{http://www.theory.caltech.edu/~preskill/ph219/topological.ps}.

\bibitem[{\citenamefont{Moore and Read}(1991)}]{MR9162}
\bibinfo{author}{\bibfnamefont{G.}~\bibnamefont{Moore}} \bibnamefont{and}
  \bibinfo{author}{\bibfnamefont{N.}~\bibnamefont{Read}},
  \bibinfo{journal}{Nucl. Phys. B} \textbf{\bibinfo{volume}{360}},
  \bibinfo{pages}{362} (\bibinfo{year}{1991}).

\bibitem[{\citenamefont{Read and Rezayi}(1999)}]{RR9984}
\bibinfo{author}{\bibfnamefont{N.}~\bibnamefont{Read}} \bibnamefont{and}
  \bibinfo{author}{\bibfnamefont{E.}~\bibnamefont{Rezayi}},
  \bibinfo{journal}{Phys. Rev. B} \textbf{\bibinfo{volume}{59}},
  \bibinfo{pages}{8084} (\bibinfo{year}{1999}).

\bibitem[{\citenamefont{Wen and Wang}(2008{\natexlab{a}})}]{WW0808}
\bibinfo{author}{\bibfnamefont{X.-G.} \bibnamefont{Wen}} \bibnamefont{and}
  \bibinfo{author}{\bibfnamefont{Z.}~\bibnamefont{Wang}},
  \bibinfo{journal}{Physical Review B} \textbf{\bibinfo{volume}{77}},
  \bibinfo{pages}{235108} (\bibinfo{year}{2008}{\natexlab{a}}).

\bibitem[{\citenamefont{Wen and Wang}(2008{\natexlab{b}})}]{WW0809}
\bibinfo{author}{\bibfnamefont{X.-G.} \bibnamefont{Wen}} \bibnamefont{and}
  \bibinfo{author}{\bibfnamefont{Z.}~\bibnamefont{Wang}},
  \bibinfo{journal}{Physical Review B} \textbf{\bibinfo{volume}{78}},
  \bibinfo{pages}{155109} (\bibinfo{year}{2008}{\natexlab{b}}).

\bibitem[{\citenamefont{Barkeshli and Wen}(2009{\natexlab{a}})}]{BW0932}
\bibinfo{author}{\bibfnamefont{M.}~\bibnamefont{Barkeshli}} \bibnamefont{and}
  \bibinfo{author}{\bibfnamefont{X.-G.} \bibnamefont{Wen}},
  \bibinfo{journal}{Phys. Rev. B} \textbf{\bibinfo{volume}{79}},
  \bibinfo{pages}{195132} (\bibinfo{year}{2009}{\natexlab{a}}).

\bibitem[{\citenamefont{Barkeshli and Wen}(2009{\natexlab{b}})}]{BW0937}
\bibinfo{author}{\bibfnamefont{M.}~\bibnamefont{Barkeshli}} \bibnamefont{and}
  \bibinfo{author}{\bibfnamefont{X.-G.} \bibnamefont{Wen}}
  (\bibinfo{year}{2009}{\natexlab{b}}), \eprint{arXiv:0906.0337}.

\bibitem[{\citenamefont{Lu et~al.}(2010)\citenamefont{Lu, Wen, Wang, and
  Wang}}]{LW1024}
\bibinfo{author}{\bibfnamefont{Y.-M.} \bibnamefont{Lu}},
  \bibinfo{author}{\bibfnamefont{X.-G.} \bibnamefont{Wen}},
  \bibinfo{author}{\bibfnamefont{Z.}~\bibnamefont{Wang}}, \bibnamefont{and}
  \bibinfo{author}{\bibfnamefont{Z.}~\bibnamefont{Wang}},
  \bibinfo{journal}{Phys. Rev. B} \textbf{\bibinfo{volume}{81}},
  \bibinfo{pages}{115124} (\bibinfo{year}{2010}).

\bibitem[{\citenamefont{Kitaev}(2003)}]{K032}
\bibinfo{author}{\bibfnamefont{A.~Y.} \bibnamefont{Kitaev}},
  \bibinfo{journal}{Ann. Phys. (N.Y.)} \textbf{\bibinfo{volume}{303}},
  \bibinfo{pages}{2} (\bibinfo{year}{2003}).

\bibitem[{\citenamefont{Freedman et~al.}(2003)\citenamefont{Freedman, Kitaev,
  Larsen, and Wang}}]{FKL0331}
\bibinfo{author}{\bibfnamefont{M.~H.} \bibnamefont{Freedman}},
  \bibinfo{author}{\bibfnamefont{A.}~\bibnamefont{Kitaev}},
  \bibinfo{author}{\bibfnamefont{M.~J.} \bibnamefont{Larsen}},
  \bibnamefont{and} \bibinfo{author}{\bibfnamefont{Z.}~\bibnamefont{Wang}},
  \bibinfo{journal}{Bull. Amer. Math. Soc.} \textbf{\bibinfo{volume}{40}},
  \bibinfo{pages}{31} (\bibinfo{year}{2003}).

\bibitem[{\citenamefont{Dennis et~al.}(2002)\citenamefont{Dennis, Kitaev,
  Landahl, and Preskill}}]{DKL0252}
\bibinfo{author}{\bibfnamefont{E.}~\bibnamefont{Dennis}},
  \bibinfo{author}{\bibfnamefont{A.}~\bibnamefont{Kitaev}},
  \bibinfo{author}{\bibfnamefont{A.}~\bibnamefont{Landahl}}, \bibnamefont{and}
  \bibinfo{author}{\bibfnamefont{J.}~\bibnamefont{Preskill}},
  \bibinfo{journal}{J. Math. Phys.} \textbf{\bibinfo{volume}{43}},
  \bibinfo{pages}{4452} (\bibinfo{year}{2002}).

\bibitem[{\citenamefont{Halperin}(1983)}]{H8375}
\bibinfo{author}{\bibfnamefont{B.}~\bibnamefont{Halperin}},
  \bibinfo{journal}{Helvetica Physica Acta} \textbf{\bibinfo{volume}{56}},
  \bibinfo{pages}{75} (\bibinfo{year}{1983}).

\bibitem[{\citenamefont{Barkeshli and Wen}(2010{\natexlab{a}})}]{BW102}
\bibinfo{author}{\bibfnamefont{M.}~\bibnamefont{Barkeshli}} \bibnamefont{and}
  \bibinfo{author}{\bibfnamefont{X.-G.} \bibnamefont{Wen}}
  (\bibinfo{year}{2010}{\natexlab{a}}), \eprint{arXiv:1007.2030}.

\bibitem[{\citenamefont{Wen}(2000)}]{W0050}
\bibinfo{author}{\bibfnamefont{X.-G.} \bibnamefont{Wen}},
  \bibinfo{journal}{Phys. Rev. Lett.} \textbf{\bibinfo{volume}{84}},
  \bibinfo{pages}{3950} (\bibinfo{year}{2000}).

\bibitem[{\citenamefont{Read and Green}(2000)}]{RG0067}
\bibinfo{author}{\bibfnamefont{N.}~\bibnamefont{Read}} \bibnamefont{and}
  \bibinfo{author}{\bibfnamefont{D.}~\bibnamefont{Green}},
  \bibinfo{journal}{Phys. Rev. B} \textbf{\bibinfo{volume}{61}},
  \bibinfo{pages}{10267} (\bibinfo{year}{2000}).

\bibitem[{\citenamefont{Bais and Slingerland}(2009)}]{BS0916}
\bibinfo{author}{\bibfnamefont{F.}~\bibnamefont{Bais}} \bibnamefont{and}
  \bibinfo{author}{\bibfnamefont{J.}~\bibnamefont{Slingerland}},
  \bibinfo{journal}{Phys. Rev. B} \textbf{\bibinfo{volume}{79}},
  \bibinfo{pages}{045316} (\bibinfo{year}{2009}).

\bibitem[{\citenamefont{Barkeshli and Wen}(2010{\natexlab{b}})}]{BW1023}
\bibinfo{author}{\bibfnamefont{M.}~\bibnamefont{Barkeshli}} \bibnamefont{and}
  \bibinfo{author}{\bibfnamefont{X.-G.} \bibnamefont{Wen}},
  \bibinfo{journal}{Phys. Rev. B} \textbf{\bibinfo{volume}{81}},
  \bibinfo{pages}{045323} (\bibinfo{year}{2010}{\natexlab{b}}).

\bibitem[{\citenamefont{Blok and Wen}(1990)}]{BW9045}
\bibinfo{author}{\bibfnamefont{B.}~\bibnamefont{Blok}} \bibnamefont{and}
  \bibinfo{author}{\bibfnamefont{X.~G.} \bibnamefont{Wen}},
  \bibinfo{journal}{Phys. Rev. B} \textbf{\bibinfo{volume}{42}},
  \bibinfo{pages}{8145} (\bibinfo{year}{1990}).

\bibitem[{\citenamefont{Wen}(1995)}]{Wtoprev}
\bibinfo{author}{\bibfnamefont{X.-G.} \bibnamefont{Wen}},
  \bibinfo{journal}{Advances in Physics} \textbf{\bibinfo{volume}{44}},
  \bibinfo{pages}{405} (\bibinfo{year}{1995}).

\bibitem[{\citenamefont{Verlinde}(1988)}]{V8860}
\bibinfo{author}{\bibfnamefont{E.}~\bibnamefont{Verlinde}},
  \bibinfo{journal}{Nucl. Phys. B} \textbf{\bibinfo{volume}{300}},
  \bibinfo{pages}{360} (\bibinfo{year}{1988}).

\bibitem[{\citenamefont{Senthil and Fisher}(2000)}]{SF0050}
\bibinfo{author}{\bibfnamefont{T.}~\bibnamefont{Senthil}} \bibnamefont{and}
  \bibinfo{author}{\bibfnamefont{M.~P.~A.} \bibnamefont{Fisher}},
  \bibinfo{journal}{Phys. Rev. B} \textbf{\bibinfo{volume}{62}},
  \bibinfo{pages}{7850} (\bibinfo{year}{2000}).

\bibitem[{\citenamefont{Wen}(1991)}]{Wnab}
\bibinfo{author}{\bibfnamefont{X.-G.} \bibnamefont{Wen}},
  \bibinfo{journal}{Phys. Rev. Lett.} \textbf{\bibinfo{volume}{66}},
  \bibinfo{pages}{802} (\bibinfo{year}{1991}).

\bibitem[{\citenamefont{Wen}(1999)}]{W9927}
\bibinfo{author}{\bibfnamefont{X.-G.} \bibnamefont{Wen}},
  \bibinfo{journal}{Phys. Rev. B} \textbf{\bibinfo{volume}{60}},
  \bibinfo{pages}{8827} (\bibinfo{year}{1999}).

\bibitem[{\citenamefont{Wen}(1992)}]{W9211}
\bibinfo{author}{\bibfnamefont{X.-G.} \bibnamefont{Wen}},
  \bibinfo{journal}{Int. J. Mod. Phys.} \textbf{\bibinfo{volume}{B6}},
  \bibinfo{pages}{1711} (\bibinfo{year}{1992}).

\bibitem[{\citenamefont{Wen and Zee}(1992)}]{WZ9290}
\bibinfo{author}{\bibfnamefont{X.-G.} \bibnamefont{Wen}} \bibnamefont{and}
  \bibinfo{author}{\bibfnamefont{A.}~\bibnamefont{Zee}},
  \bibinfo{journal}{Phys. Rev. B} \textbf{\bibinfo{volume}{46}},
  \bibinfo{pages}{2290} (\bibinfo{year}{1992}).

\bibitem[{\citenamefont{Witten}(1989)}]{W8951}
\bibinfo{author}{\bibfnamefont{E.}~\bibnamefont{Witten}},
  \bibinfo{journal}{Comm. Math. Phys.} \textbf{\bibinfo{volume}{121}},
  \bibinfo{pages}{351} (\bibinfo{year}{1989}).

\bibitem[{\citenamefont{Moore and Seiberg}(1989{\natexlab{a}})}]{MS8922}
\bibinfo{author}{\bibfnamefont{G.}~\bibnamefont{Moore}} \bibnamefont{and}
  \bibinfo{author}{\bibfnamefont{N.}~\bibnamefont{Seiberg}},
  \bibinfo{journal}{Phys. Lett. B} \textbf{\bibinfo{volume}{220}},
  \bibinfo{pages}{422} (\bibinfo{year}{1989}{\natexlab{a}}).

\bibitem[{\citenamefont{Dijkgraaf and Witten}(1990)}]{DW9093}
\bibinfo{author}{\bibfnamefont{R.}~\bibnamefont{Dijkgraaf}} \bibnamefont{and}
  \bibinfo{author}{\bibfnamefont{E.}~\bibnamefont{Witten}},
  \bibinfo{journal}{Communications in Mathematical Physics}
  \textbf{\bibinfo{volume}{129}}, \bibinfo{pages}{393} (\bibinfo{year}{1990}).

\bibitem[{\citenamefont{Dijkgraaf et~al.}(1989)\citenamefont{Dijkgraaf, Vafa,
  Verlinde, and Verlinde}}]{DV8985}
\bibinfo{author}{\bibfnamefont{R.}~\bibnamefont{Dijkgraaf}},
  \bibinfo{author}{\bibfnamefont{C.}~\bibnamefont{Vafa}},
  \bibinfo{author}{\bibfnamefont{E.}~\bibnamefont{Verlinde}}, \bibnamefont{and}
  \bibinfo{author}{\bibfnamefont{H.}~\bibnamefont{Verlinde}},
  \bibinfo{journal}{Comm. Math. Phys.} \textbf{\bibinfo{volume}{123}},
  \bibinfo{pages}{485} (\bibinfo{year}{1989}).

\bibitem[{\citenamefont{Francesco et~al.}(1997)\citenamefont{Francesco,
  Mathieu, and Senechal}}]{FMCFT}
\bibinfo{author}{\bibfnamefont{P.~D.} \bibnamefont{Francesco}},
  \bibinfo{author}{\bibfnamefont{P.}~\bibnamefont{Mathieu}}, \bibnamefont{and}
  \bibinfo{author}{\bibfnamefont{D.}~\bibnamefont{Senechal}},
  \emph{\bibinfo{title}{Conformal Field Theory}}
  (\bibinfo{publisher}{Springer}, \bibinfo{year}{1997}).

\bibitem[{\citenamefont{Bernevig and Haldane}(2008)}]{BH0802}
\bibinfo{author}{\bibfnamefont{B.~A.} \bibnamefont{Bernevig}} \bibnamefont{and}
  \bibinfo{author}{\bibfnamefont{F.~D.~M.} \bibnamefont{Haldane}},
  \bibinfo{journal}{Phys. Rev. Lett.} \textbf{\bibinfo{volume}{100}},
  \bibinfo{pages}{246802} (\bibinfo{year}{2008}).

\bibitem[{\citenamefont{Bergholtz and Karlhede}(2006)}]{BK0601}
\bibinfo{author}{\bibfnamefont{E.~J.} \bibnamefont{Bergholtz}}
  \bibnamefont{and} \bibinfo{author}{\bibfnamefont{A.}~\bibnamefont{Karlhede}},
  \bibinfo{journal}{Journal of Statistical Mechanics: Theory and Experiment}
  \textbf{\bibinfo{volume}{2006}}, \bibinfo{pages}{L04001}
  (\bibinfo{year}{2006}).

\bibitem[{\citenamefont{Seidel and Lee}(2006)}]{SL0604}
\bibinfo{author}{\bibfnamefont{A.}~\bibnamefont{Seidel}} \bibnamefont{and}
  \bibinfo{author}{\bibfnamefont{D.-H.} \bibnamefont{Lee}},
  \bibinfo{journal}{Phys. Rev. Lett.} \textbf{\bibinfo{volume}{97}},
  \bibinfo{pages}{056804} (\bibinfo{year}{2006}).

\bibitem[{\citenamefont{Fradkin and Shenker}(1979)}]{FS7982}
\bibinfo{author}{\bibfnamefont{E.}~\bibnamefont{Fradkin}} \bibnamefont{and}
  \bibinfo{author}{\bibfnamefont{S.~H.} \bibnamefont{Shenker}},
  \bibinfo{journal}{Phys. Rev. D} \textbf{\bibinfo{volume}{19}},
  \bibinfo{pages}{3682} (\bibinfo{year}{1979}).

\bibitem[{\citenamefont{Zamolodchikov and Fateev}(1985)}]{ZF8515}
\bibinfo{author}{\bibfnamefont{A.}~\bibnamefont{Zamolodchikov}}
  \bibnamefont{and} \bibinfo{author}{\bibfnamefont{V.}~\bibnamefont{Fateev}},
  \bibinfo{journal}{Sov. Phys. JETP} \textbf{\bibinfo{volume}{62}},
  \bibinfo{pages}{215} (\bibinfo{year}{1985}).

\bibitem[{\citenamefont{Read}(2009)}]{R0908}
\bibinfo{author}{\bibfnamefont{N.}~\bibnamefont{Read}}, \bibinfo{journal}{Phys.
  Rev. B} \textbf{\bibinfo{volume}{79}}, \bibinfo{pages}{045308}
  (\bibinfo{year}{2009}).

\bibitem[{\citenamefont{Simon et~al.}(2007)\citenamefont{Simon, Rezayi, and
  Cooper}}]{SR0718}
\bibinfo{author}{\bibfnamefont{S.~H.} \bibnamefont{Simon}},
  \bibinfo{author}{\bibfnamefont{E.~H.} \bibnamefont{Rezayi}},
  \bibnamefont{and} \bibinfo{author}{\bibfnamefont{N.~R.}
  \bibnamefont{Cooper}}, \bibinfo{journal}{Phys. Rev. B}
  \textbf{\bibinfo{volume}{75}}, \bibinfo{pages}{075318}
  (\bibinfo{year}{2007}).

\bibitem[{\citenamefont{Ardonne and Schoutens}(1999)}]{AS9996}
\bibinfo{author}{\bibfnamefont{E.}~\bibnamefont{Ardonne}} \bibnamefont{and}
  \bibinfo{author}{\bibfnamefont{K.}~\bibnamefont{Schoutens}},
  \bibinfo{journal}{Physical Review Letters} \textbf{\bibinfo{volume}{82}},
  \bibinfo{pages}{5096} (\bibinfo{year}{1999}).

\bibitem[{\citenamefont{Ardonne et~al.}(2002)\citenamefont{Ardonne, van
  Lankvelt, Ludwig, and Schoutens}}]{AL0205}
\bibinfo{author}{\bibfnamefont{E.}~\bibnamefont{Ardonne}},
  \bibinfo{author}{\bibfnamefont{F.}~\bibnamefont{van Lankvelt}},
  \bibinfo{author}{\bibfnamefont{A.}~\bibnamefont{Ludwig}}, \bibnamefont{and}
  \bibinfo{author}{\bibfnamefont{K.}~\bibnamefont{Schoutens}},
  \bibinfo{journal}{Physical Review B} \textbf{\bibinfo{volume}{65}}
  (\bibinfo{year}{2002}).

\bibitem[{\citenamefont{Moore and Seiberg}(1989{\natexlab{b}})}]{MS8916}
\bibinfo{author}{\bibfnamefont{G.}~\bibnamefont{Moore}} \bibnamefont{and}
  \bibinfo{author}{\bibfnamefont{N.}~\bibnamefont{Seiberg}},
  \bibinfo{journal}{Nucl. Phys. B} \textbf{\bibinfo{volume}{313}},
  \bibinfo{pages}{16} (\bibinfo{year}{1989}{\natexlab{b}}).

\bibitem[{\citenamefont{Green}(2001)}]{G01}
\bibinfo{author}{\bibfnamefont{D.}~\bibnamefont{Green}}, Ph.D. thesis,
  \bibinfo{school}{Yale University} (\bibinfo{year}{2001}).

\bibitem[{\citenamefont{Lay et~al.}(1997)\citenamefont{Lay, Jungwirth, Smrcka,
  and Shayegan}}]{LJS9792}
\bibinfo{author}{\bibfnamefont{T.~S.} \bibnamefont{Lay}},
  \bibinfo{author}{\bibfnamefont{T.}~\bibnamefont{Jungwirth}},
  \bibinfo{author}{\bibfnamefont{L.}~\bibnamefont{Smrcka}}, \bibnamefont{and}
  \bibinfo{author}{\bibfnamefont{M.}~\bibnamefont{Shayegan}},
  \bibinfo{journal}{Phys. Rev. B} \textbf{\bibinfo{volume}{56}},
  \bibinfo{pages}{R7092} (\bibinfo{year}{1997}).

\bibitem[{\citenamefont{Suen et~al.}(1994)\citenamefont{Suen, Manoharan, Ying,
  Santos, and Shayegan}}]{SM9405}
\bibinfo{author}{\bibfnamefont{Y.~W.} \bibnamefont{Suen}},
  \bibinfo{author}{\bibfnamefont{H.~C.} \bibnamefont{Manoharan}},
  \bibinfo{author}{\bibfnamefont{X.}~\bibnamefont{Ying}},
  \bibinfo{author}{\bibfnamefont{M.~B.} \bibnamefont{Santos}},
  \bibnamefont{and} \bibinfo{author}{\bibfnamefont{M.}~\bibnamefont{Shayegan}},
  \bibinfo{journal}{Phys. Rev. Lett.} \textbf{\bibinfo{volume}{72}},
  \bibinfo{pages}{3405} (\bibinfo{year}{1994}).

\bibitem[{\citenamefont{Willett et~al.}(1993)\citenamefont{Willett, Ruel,
  Paalanen, West, and Pfeiffer}}]{WR9344}
\bibinfo{author}{\bibfnamefont{R.~L.} \bibnamefont{Willett}},
  \bibinfo{author}{\bibfnamefont{R.~R.} \bibnamefont{Ruel}},
  \bibinfo{author}{\bibfnamefont{M.~A.} \bibnamefont{Paalanen}},
  \bibinfo{author}{\bibfnamefont{K.~W.} \bibnamefont{West}}, \bibnamefont{and}
  \bibinfo{author}{\bibfnamefont{L.~N.} \bibnamefont{Pfeiffer}},
  \bibinfo{journal}{Phys. Rev. B} \textbf{\bibinfo{volume}{47}},
  \bibinfo{pages}{7344} (\bibinfo{year}{1993}).

\bibitem[{\citenamefont{Xia et~al.}(2004)\citenamefont{Xia, Pan, Vicente,
  Adams, Sullivan, Stormer, Tsui, Pfeiffer, Baldwin, and West}}]{XP0409}
\bibinfo{author}{\bibfnamefont{J.~S.} \bibnamefont{Xia}},
  \bibinfo{author}{\bibfnamefont{W.}~\bibnamefont{Pan}},
  \bibinfo{author}{\bibfnamefont{C.~L.} \bibnamefont{Vicente}},
  \bibinfo{author}{\bibfnamefont{E.~D.} \bibnamefont{Adams}},
  \bibinfo{author}{\bibfnamefont{N.~S.} \bibnamefont{Sullivan}},
  \bibinfo{author}{\bibfnamefont{H.~L.} \bibnamefont{Stormer}},
  \bibinfo{author}{\bibfnamefont{D.~C.} \bibnamefont{Tsui}},
  \bibinfo{author}{\bibfnamefont{L.~N.} \bibnamefont{Pfeiffer}},
  \bibinfo{author}{\bibfnamefont{K.~W.} \bibnamefont{Baldwin}},
  \bibnamefont{and} \bibinfo{author}{\bibfnamefont{K.~W.} \bibnamefont{West}},
  \bibinfo{journal}{Phys. Rev. Lett.} \textbf{\bibinfo{volume}{93}},
  \bibinfo{pages}{176809} (\bibinfo{year}{2004}).

\bibitem[{\citenamefont{Choi et~al.}(2008)\citenamefont{Choi, Kang, Das~Sarma,
  Pfeiffer, and West}}]{CK0801}
\bibinfo{author}{\bibfnamefont{H.~C.} \bibnamefont{Choi}},
  \bibinfo{author}{\bibfnamefont{W.}~\bibnamefont{Kang}},
  \bibinfo{author}{\bibfnamefont{S.}~\bibnamefont{Das~Sarma}},
  \bibinfo{author}{\bibfnamefont{L.~N.} \bibnamefont{Pfeiffer}},
  \bibnamefont{and} \bibinfo{author}{\bibfnamefont{K.~W.} \bibnamefont{West}},
  \bibinfo{journal}{Phys. Rev. B} \textbf{\bibinfo{volume}{77}},
  \bibinfo{pages}{081301(R)} (\bibinfo{year}{2008}).

\bibitem[{\citenamefont{Rezayi and Read}(2009)}]{RR0906}
\bibinfo{author}{\bibfnamefont{E.~H.} \bibnamefont{Rezayi}} \bibnamefont{and}
  \bibinfo{author}{\bibfnamefont{N.}~\bibnamefont{Read}},
  \bibinfo{journal}{Phys. Rev. B} \textbf{\bibinfo{volume}{79}},
  \bibinfo{pages}{075306} (\bibinfo{year}{2009}).

\bibitem[{\citenamefont{Bonderson and Slingerland}(2008)}]{BS0823}
\bibinfo{author}{\bibfnamefont{P.}~\bibnamefont{Bonderson}} \bibnamefont{and}
  \bibinfo{author}{\bibfnamefont{J.~K.} \bibnamefont{Slingerland}},
  \bibinfo{journal}{Phys. Rev. B} \textbf{\bibinfo{volume}{78}},
  \bibinfo{pages}{125323} (\bibinfo{year}{2008}).

\bibitem[{\citenamefont{Bonderson et~al.}(2009)\citenamefont{Bonderson,
  Feiguin, Moller, and Slingerland}}]{BF0964}
\bibinfo{author}{\bibfnamefont{P.}~\bibnamefont{Bonderson}},
  \bibinfo{author}{\bibfnamefont{A.~E.} \bibnamefont{Feiguin}},
  \bibinfo{author}{\bibfnamefont{G.}~\bibnamefont{Moller}}, \bibnamefont{and}
  \bibinfo{author}{\bibfnamefont{J.}~\bibnamefont{Slingerland}},
  \bibinfo{journal}{arXiv:} \textbf{\bibinfo{volume}{0901}},
  \bibinfo{pages}{4965} (\bibinfo{year}{2009}).

\bibitem[{\citenamefont{Rowell and Wang}(2010)}]{RW1041}
\bibinfo{author}{\bibfnamefont{E.~C.} \bibnamefont{Rowell}} \bibnamefont{and}
  \bibinfo{author}{\bibfnamefont{Z.}~\bibnamefont{Wang}}
  (\bibinfo{year}{2010}), \eprint{arXiv:1009.0241}.

\bibitem[{\citenamefont{Gu et~al.}(2010)\citenamefont{Gu, Wang, and
  Wen}}]{GW1017}
\bibinfo{author}{\bibfnamefont{Z.-C.} \bibnamefont{Gu}},
  \bibinfo{author}{\bibfnamefont{Z.}~\bibnamefont{Wang}}, \bibnamefont{and}
  \bibinfo{author}{\bibfnamefont{X.-G.} \bibnamefont{Wen}}
  (\bibinfo{year}{2010}), \eprint{arXiv:1010.1517}.

\end{thebibliography}

\end{document}